%% file: main.tex
\documentclass[11pt,a4paper]{article}

\usepackage[T1]{fontenc}
\usepackage[utf8]{inputenc}
\IfFileExists{lmodern.sty}{\usepackage{lmodern}}{}   
\usepackage[a4paper,top=2.6cm,bottom=2.6cm,left=2.7cm,right=2.7cm]{geometry}

\usepackage{amsmath,amssymb,amsfonts}
\usepackage{booktabs}
\usepackage{array}
\usepackage{graphicx}
\usepackage{longtable}
\usepackage{calc}          
\IfFileExists{lmodern.sty}{\usepackage{microtype}}{\usepackage[expansion=false]{microtype}}
\usepackage{setspace}
\usepackage[authoryear,round]{natbib}
\usepackage[font=small,labelfont=bf,skip=6pt]{caption}
\usepackage{titlesec}
\usepackage[hidelinks,breaklinks]{hyperref}

\graphicspath{{figures/}}
\setcitestyle{aysep{}}

\titleformat{\section}{\normalfont\large\bfseries}{\thesection}{0.8em}{}
\titleformat{\subsection}{\normalfont\normalsize\bfseries}{\thesubsection}{0.8em}{}
\titlespacing*{\section}{0pt}{2.2ex plus 1ex minus .2ex}{1.2ex plus .2ex}
\titlespacing*{\subsection}{0pt}{1.8ex plus 1ex minus .2ex}{1.0ex plus .2ex}

\providecommand{\tightlist}{\setlength{\itemsep}{0pt}\setlength{\parskip}{0pt}}

\title{\bfseries The Reconfiguration Premium:\\[2pt]
       \large Co-movement Structure as an Unspanned Dimension of the\\
       Variance Risk Premium}
\author{%
  Lucas Carvalho\thanks{%
    ISEG --- Instituto Superior de Economia e Gest\~ao, Universidade de Lisboa,
    Lisbon, Portugal. Email: \texttt{lucas\_carvalho@phd.iseg.ulisboa.pt}.
    This research received no specific grant from any funding agency.
    Replication code and data:
    \url{https://github.com/lucas-p-carvalho/rec-anatomy}.
    \emph{Generative-AI disclosure:} analysis code and manuscript drafting were
    carried out with the assistance of Claude (Anthropic), using the Opus
    models (versions 4.7 through 5), Sonnet and Fable, to implement and run
    the analysis pipeline, to
    generate figures from its output, and to draft text from the author's
    specifications. The research design, hypotheses, pre-registered null tests
    and all interpretive decisions are the author's. Every quantity reported was
    regenerated from source data and verified against the analysis logs, and the
    bibliography was audited in both directions. The author takes full
    responsibility for all content, and all remaining errors are his own.}%
}
\date{\today}

\begin{document}
\maketitle

\begin{abstract}
\noindent
\input{abstract}
\vspace{6pt}

\noindent\textbf{Keywords:} \input{keywords}
\end{abstract}

\vspace{4pt}

\input{sections/01_introduction}
\input{sections/02_data}
\input{sections/03_index}
\input{sections/04_object}
\input{sections/05_priced}
\input{sections/06_mechanism}
\input{sections/07_boundaries}
\input{sections/08_discussion}

\appendix
\counterwithin{table}{section}
\renewcommand{\thetable}{\thesection\arabic{table}}
\input{appendices/A_inference}
\input{appendices/B_robustness}
\input{appendices/C_harvest}
\input{appendices/D_panel}

\input{sections/09_prior_work}

\bibliography{references}

\end{document}

%% file: abstract.tex
Hedge ratios, factor models and diversified portfolios all rest on an estimate of which firms move together. That estimate is not stable: firms migrate between the groupings the market treats as coherent, and when enough migrate the organizing axes of the cross-section turn. We measure the rate of that turning as the mean squared sine of the principal angles between subdominant eigenspaces of consecutive twelve-month S\&P 500 correlation matrices. A typical month rewrites a fifth of the structure and carries four-fifths forward. That rate is priced: it couples to the aggregate variance risk premium at \(t = 5.40\), no level measure correlates above 0.32, and the implied-correlation surface spans at most 6.7 percent of it. Only the persistent component is priced --- the premium compensates the pace of revision, not the distance traveled. The mechanism is prepayment: implied variance rises on impact, volatility follows two to three quarters later (simulated-null \(p < 0.03\) at \(h = 1\)--9), and the premium converges as it arrives. Three pre-registered boundaries hold: no timing alpha, no crash protection, and a downside version inseparable from intensity. The premium is, in part, rent on exposure held over a map still being redrawn.

%% file: keywords.tex
variance risk premium, correlation structure, principal angles, eigenvector rotation, unspanned state variables, option-implied correlation.

%% file: sections/01_introduction.tex
\section{Introduction}\label{sec:intro}

Every risk instrument in use assumes a map of the market. A hedge ratio is a claim about which two assets move together; a factor model is a partition of firms by what drives them; a diversified book is a bet that its holdings sit in different regions of that partition. All three are estimated from the correlation matrix, and all three are estimated as though the map itself were fixed and only the distances on it varied.

It is not fixed. Firms migrate between the groupings the market treats as coherent --- a utility priced as a bond substitute in one period is priced as an industrial in another --- and when enough of them migrate together, the axes along which the cross-section is organized turn. A hedge estimated before the turn is a claim about a structure that no longer holds. This paper measures the rate at which that structure is rewritten, and asks what the rate costs. The answer is that it is expensive. Option-implied variance rises the moment the rewriting accelerates; the volatility that justifies the higher price arrives two to three quarters later; and no traded instrument spans the state variable, so the compensation cannot be arbitraged into an existing contract. Part of what the variance risk premium pays for is the cost of holding exposure over a map that is still being redrawn.

The measurement is available in the object the literature already estimates. Let \(C_t\) denote the correlation matrix of equity returns estimated on a window ending at \(t\), with spectral decomposition

\begin{equation}\label{eq:spec}
C_t = V_t \Lambda_t V_t^{\prime}, \qquad \Lambda_t = \mathrm{diag}(\lambda_{1,t}, \dots, \lambda_{N,t}).
\end{equation}

Every measure of co-movement that the asset-pricing literature currently prices is a functional of \(\Lambda_t\) alone. Index variance is \(w^{\prime} \Sigma_t w\), a quadratic form dominated by the leading eigenvalue; average pairwise correlation is an affine function of \(\iota^{\prime} C_t \iota\); option-implied correlation is a ratio of index to constituent implied variances, hence again a level. The absorption ratio of \citet{kritzman2011pca} is the share of variance in the leading eigenvalues. None of these objects is sensitive to the orientation of \(V_t\): they are invariant to any rotation of the eigenbasis that leaves the spectrum fixed. The residual dimension of \eqref{eq:spec} is therefore \(\dot V_t\), the motion of the eigenbasis itself, and it has not been priced. The decomposition separates two independent kinds of information, and the literature has priced only one of them.

\begin{longtable}[]{@{}
  >{\raggedright\arraybackslash}p{(\columnwidth - 4\tabcolsep) * \real{0.3333}}
  >{\raggedright\arraybackslash}p{(\columnwidth - 4\tabcolsep) * \real{0.3333}}
  >{\raggedright\arraybackslash}p{(\columnwidth - 4\tabcolsep) * \real{0.3333}}@{}}
\toprule\noalign{}
\begin{minipage}[b]{\linewidth}\raggedright
\end{minipage} & \begin{minipage}[b]{\linewidth}\raggedright
What it carries
\end{minipage} & \begin{minipage}[b]{\linewidth}\raggedright
Priced?
\end{minipage} \\
\midrule\noalign{}
\endhead
\bottomrule\noalign{}
\endlastfoot
\(\Lambda_t\), the eigenvalues & \emph{how much} firms co-move & Yes --- index variance, correlation level, option-implied correlation, the absorption ratio \\
\(V_t\), the eigenvectors & \emph{which} firms co-move --- the grouping & Never \\
\end{longtable}

This paper prices the motion of the second.

The gap this leaves is sharp. On one shore, the \emph{level} of correlation is known to carry a risk premium. \citet{driessen2009price} show that index options are expensive relative to a portfolio of individual options precisely because correlation risk is priced; \citet{buss2019expected} show that option-implied correlation predicts market returns as a procyclical state variable; \citet{buraschi2014} find exposure to correlation risk priced in the cross-section of hedge fund returns; \citet{carr2009variance} and \citet{bollerslev2009expected} document the aggregate variance risk premium and its predictive content. On the other shore, the \emph{structure} of correlation has been described for three decades without reference to prices. \citet{laloux1999noise} and \citet{plerou2002rmt} established which part of the spectrum is signal; \citet{onnela2003dynamics} traced the deformation of the minimum spanning tree through crises; the stochastic-geometry lineage of \citet{vilelamendes2002process}, \citet{vilelamendes2003reconstructing}, \citet{araujo2007geometry} and \citet{carvalho2024etf} reconstructed an economic metric space from returns and tracked its dimensional collapse around crashes. Closest to the construction used here, \citet{eleuterio2014portfolios} showed that the small-eigenvalue directions of that metric space carry economic content rather than noise, the best-performing portfolios loading on non-dominant axes --- the precedent for working in the subdominant block at all. That literature measures geometry; it does not ask what geometry costs.

This paper asks. We construct an index of subdominant eigenspace rotation, establish that it is distinct from every level measure and largely unspanned by the traded implied-correlation surface, show that it couples robustly to the aggregate variance risk premium, and then decompose the coupling along three axes: which component of rotation is priced, through what temporal mechanism, and what the coupling is not.

The narrowing proceeds one further step. Write the panel in approximate factor form, \(r_t = B_t f_t + \varepsilon_t\), so that \(C_t\) is generated by loadings \(B_t\) and factor covariance \(\Sigma_{f,t}\). Eigenvector rotation has two sources. Time-varying factor volatilities rotate the eigenbasis even with loadings held fixed, because the ordering and mixture of modes follows the relative volatilities of the underlying factors; drift in \(B_t\) rotates it for a second and distinct reason. Section~\ref{sec:discussion} shows that the priced rotation is not the first channel --- the cross-sectional churn of the volatility hierarchy is orthogonal to the index and unpriced --- and not discrete substitution of one economic axis for another, since labeled mode-switching does not absorb the pricing. By elimination within the \(B \cdot f\) decomposition, the priced object is \(\dot B_t\): continuous, diffuse revision of which firms load on the market's operative classification axes.

Six results follow.

\begin{itemize}
\tightlist
\item
  \emph{(i) The estimator is sound.} Modes two and three sit above the refitted Marchenko--Pastur edge in essentially every window; the Davis--Kahan degradation channel is real but bounds at 12.1\% of the index's variance; the coupling is unchanged with the entire mechanical model as controls. \textbf{Meaning: the index measures genuine structural rotation, not eigenvalue-crowding noise.}
\item
  \emph{(ii) The index is a distinct object.} No level measure correlates with it above 0.32; the implied-correlation surface spans 5.0--6.7\% of it. \textbf{Meaning: it is new information relative to everything the options market already trades.}
\item
  \emph{(iii) The priced component is the persistent one.} The three-month average carries the entire coupling (\emph{t} = 5.40), the monthly innovation none (\emph{t} = 0.30); one-month rotation drives out twelve-month drift (4.76 vs 1.38). \textbf{Meaning: the market prices sustained ongoing revision --- the pace, not the distance traveled.}
\item
  \emph{(iv) The mechanism is prepayment.} Rotation forecasts the broad volatility environment one to three quarters ahead (simulated null cleared at every \emph{h} = 1--9 under both lag specifications; \emph{h} = 10 under the primary design only); implied variance responds on impact; the premium then compresses as the forecast volatility arrives. \textbf{Meaning: the widening is payment in advance for turbulence that genuinely comes.}
\item
  \emph{(v) The option market reacts without embedding.} High rotation inverts the implied-correlation term structure (\emph{t} = $-$3.46) though the curve's level spans almost none of the index. \textbf{Meaning: option prices respond to a state variable no traded instrument contains.}
\item
  \emph{(vi) The pre-registered boundaries hold.} No timing alpha, no crash protection, and the downside version is inseparable from downside intensity at monthly resolution. \textbf{Meaning: this is attribution and state measurement, not a strategy --- and the tail story is not identifiable at this frequency.}
\end{itemize}

The contribution is therefore fivefold: the index and its validity battery; the existence result (coupling, orthogonality, spanning); the anatomy (frequency and horizon); the mechanism with simulation-based inference; and the boundaries, including one closed negative result. Section~\ref{sec:data} sets out the data and the premium convention, whose non-standard construction is disclosed as a scope condition rather than a footnote. Section~\ref{sec:index} constructs the index and establishes its validity. Section~\ref{sec:object} establishes the object. Section~\ref{sec:priced} identifies the priced component, Section~\ref{sec:mechanism} the mechanism, Section~\ref{sec:not} the boundaries. Section~\ref{sec:discussion} gives the economic magnitude, the identity of the rotating structure, and the limitations.

\begin{figure}[htbp]
\centering
\includegraphics[width=0.98\textwidth]{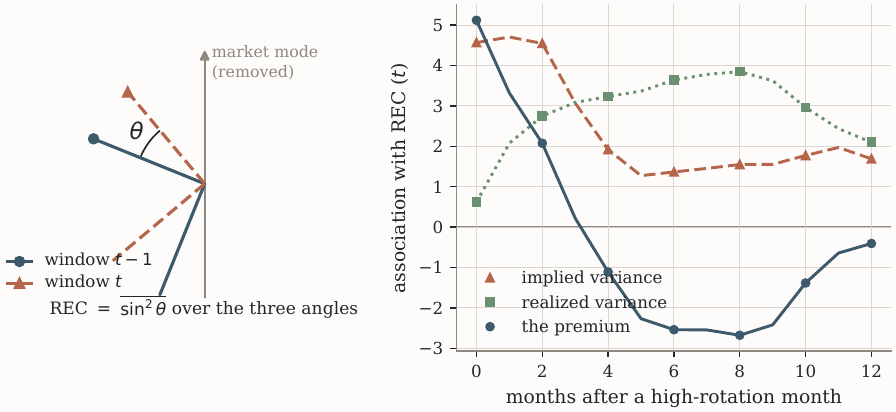}
\caption{\emph{Left:} schematic of the measurement. The market mode holds still and is removed; the subdominant frame rotates between consecutive windows, and the principal angle $\theta$ between the two positions is what the index records --- REC is the mean of $\sin^{2}\theta$ over the three angles. \emph{Right:} the two legs of the premium and their difference, as associations with rotation at horizons $h=0$ to $12$ [03b, 16]. Implied variance responds on impact; realized variance arrives two to three quarters later; the premium widens and then converges.}
\label{fig:construction}
\end{figure}

%% file: sections/02_data.tex
\section{Data and conventions}\label{sec:data}

\textbf{Panel.} The estimation panel is monthly total returns for S\&P 500 constituents including names subsequently delisted, from June 1994 to December 2025. Columns are retained if observed in at least 70 percent of months and rows if at least 80 percent of columns are observed, with the column filter reapplied after row deletion; the XLC sector is dropped because its membership is too small to label. The surviving panel is \(379 \times 430\) {[}01{]}. The thresholds trade cross-sectional breadth against window completeness: because every estimation window drops any name with a missing observation inside it, a looser column filter admits names that rarely survive a window intersection, while a tighter one shrinks the cross-section below what a stable dual-Gram estimate requires. Appendix~\ref{app:panel}.3 shows the choice is immaterial --- the coupling is significant at every threshold pair tested and the adopted pair is not the maximum. The coverage filter has a consequence that must be stated rather than discovered: it yields a near-constant-membership panel of long-history names, so the analysis universe is survivor-tilted by construction and the equal-weighted realized leg should be read as the volatility environment of the persistent large-capitalization cross-section. Appendix~\ref{app:panel} documents the composition. Each estimation window is twelve months and is restricted to names with no missing observation inside it, so the effective cross-section averages 421 names.

The sample chain is stated once and used throughout: 368 twelve-month windows yield 367 monthly rotation observations and 365 regression observations under the three-month smoothing convention. The rotation series runs from June 1995 to December 2025; its raw mean is 0.194, standard deviation 0.119, minimum 0.005 and maximum 0.526, the last attained in July 2002.

\textbf{Level measures.} Implied volatility is the month-end close of VIX (FRED series VIXCLS), the model-free implied variance of one-month S\&P 500 options \citep{whaley1993,cboe2019vix}. Realized volatility and average pairwise correlation are computed from the same panel that produces the index. Option-implied correlation is measured by the Cboe COR3M and COR1M indices (Bloomberg, 240 usable monthly observations from January 2006) and by the Cboe S\&P 500 Dispersion Index DSPX (139 observations from June 2014). Inference throughout is \citet{newey1987} with twelve lags; simulation-based inference is used wherever a forward-looking regression is run, as described in Section~\ref{sec:mechanism}.

\textbf{The premium and its scope conditions.} The primary dependent variable is

\begin{equation}\label{eq:vrp}
\mathrm{VRP}^{\log}_t \;=\; \log\!\left(\frac{\mathrm{VIX}_t}{100}\right)^{2} - \log\!\left(\frac{\mathrm{RV}_t}{100}\right)^{2},
\end{equation}

where \(\mathrm{RV}_t\) is the annualized twelve-month realized volatility of the equal-weighted portfolio of panel constituents. We verified this identification directly: the stored realized leg reproduces at correlation 1.0000 against a reconstruction of the equal-weighted portfolio's twelve-month realized volatility computed from the return panel.

Two mismatches between the legs of \eqref{eq:vrp} follow and are stated here as scope conditions rather than deferred. The legs differ in horizon, one month against twelve, and in weighting, capitalization-weighted index implied variance against equal-weighted constituent realized variance. Consequently \eqref{eq:vrp} is not the canonical variance risk premium. It is the premium of one-month implied variance over the slow, broad realized-volatility environment of the constituent cross-section, and every result below is a statement about that object. The companion paper's forward battery, rerun for this draft, confirms the boundary directly: the same smoothed index has \(|t| \leq 1.2\) against horizon-matched forward variance-swap payoffs at one, three and six months (\(-0.08\), \(+0.74\), \(+1.19\)), and \(|t| \leq 1.9\) against forward one-month capitalization-weighted index realized variance at every horizon tested. The results do not port, and are not claimed to.

The logarithmic convention in \eqref{eq:vrp} is primary and the levels premium is carried as robustness throughout (Appendix~\ref{app:robust}). The choice is itself informative. In the baseline specification of Section~\ref{sec:object} the regression \(R^2\) is 0.504 in logs against 0.149 in levels; in the four-component specification of Section~\ref{sec:priced} it is 0.526 against 0.192. The levels series is dominated by a single episode: October 2008 through December 2009 accounts for 46 percent of the total squared deviation of the levels premium about its mean. The index does not track that episode proportionally --- its own largest readings fall in March 2022, October 2018, July 2010, April 2000 and May 2020, months in which the levels premium was between \(-0.015\) and \(+0.048\) against a crisis maximum of \(+0.299\). Rotation prices the ordinary variation of the premium and not its crisis extremes, which is the pricing-space form of the no-crash result reported in Section~\ref{sec:not}.

Table~\ref{tab:desc} reports descriptive statistics for the monthly series.

\begin{table}[htbp]
\centering
\caption{Descriptive statistics, 1995:05--2025:12 ($n=368$).}
\label{tab:desc}
\begin{tabular}{@{}lrrrr@{}}
\toprule
Series & Mean & S.D. & Min & Max \\
\midrule
VIX (index points)                          & 20.08 & 7.65 & 9.51 & 59.89 \\
Realized volatility, 12m equal-weight (\%)  & 15.71 & 6.77 & 4.94 & 40.77 \\
Average pairwise correlation                & 0.252 & 0.121 & 0.043 & 0.527 \\
$\mathrm{VRP}^{\log}$                       & 0.540 & 0.697 & $-1.098$ & 2.227 \\
$\mathrm{VRP}^{\mathrm{lev}}$               & 0.017 & 0.038 & $-0.099$ & 0.299 \\
REC (raw, $n=367$)                          & 0.194 & 0.119 & 0.005 & 0.526 \\
Downside REC (raw, $n=367$)                 & 0.191 & 0.153 & --- & --- \\
\bottomrule
\end{tabular}
\end{table}

%% file: sections/03_index.tex
\section{The index: construction and validity}\label{sec:index}

\subsection{Construction}\label{sec:construction}

The construction descends from the stochastic geometry program of Vilela Mendes, Araújo and Louçã (2003) and of the authors' ETF study (Carvalho and Araújo, 2024), which identifies the low-dimensional systematic subspace of the market metric; here the analogous subdominant subspace is tracked directly in the Gram eigenstructure, with no embedding step required.

Fix a window length \(T = 12\) months. For a window ending at \(t\), let \(R_t \in \mathbb{R}^{T \times N_t}\) collect the returns of the \(N_t\) names with no missing observation inside the window, and let \(Z_t\) be \(R_t\) standardized column by column to zero mean and unit variance. Because \(T \ll N_t\), the eigenstructure is obtained in the dual space,

\begin{equation}\label{eq:gram}
G_t \;=\; \tfrac{1}{N_t} Z_t Z_t^{\prime} \;\in\; \mathbb{R}^{T \times T}, \qquad G_t = U_t D_t U_t^{\prime}, \qquad d_{1,t} \geq \dots \geq d_{T,t},
\end{equation}

which is exact: the nonzero spectrum of \(G_t\) coincides with that of \(\tfrac{1}{N_t} Z_t^{\prime} Z_t\), itself proportional to the correlation matrix \(C_t\). Stock-space loadings on the subdominant modes are recovered as

\begin{equation}\label{eq:loadings}
L_t \;=\; Z_t^{\prime} \, U_t[\,\cdot\,, 2{:}4\,] \;\in\; \mathbb{R}^{N_t \times 3}, \qquad Q_t = \mathrm{qr}(L_t),
\end{equation}

so that \(Q_t\) is an orthonormal basis for the three-dimensional subdominant eigenspace, expressed in the space of firms. The leading mode is excluded by construction. Its loading vector is near-uniform and positive in essentially every window --- it is the market factor --- and it carries no information about the \emph{classification} of firms, which is what the subdominant modes encode.

Rotation is measured between consecutive windows on the intersection of their cross-sections. Let \(\mathcal{C}_t = \mathrm{cols}(Q_{t-1}) \cap \mathrm{cols}(Q_t)\), let \(A\) and \(B\) be \(Q_{t-1}\) and \(Q_t\) restricted to \(\mathcal{C}_t\) and re-orthonormalized, and let \(\sigma_1 \geq \sigma_2 \geq \sigma_3\) be the singular values of \(A^{\prime}B\). These are the cosines of the principal angles \(\theta_i\) between the two subspaces \citep{bjorck1973angles}, and the reconfiguration index is their mean squared sine,

\begin{equation}\label{eq:rec}
\mathrm{REC}_t \;=\; \frac{1}{3}\sum_{i=1}^{3}\sin^{2}\theta_i \;=\; 1 - \frac{1}{3}\,\bigl\|A^{\prime}B\bigr\|_F^{2} \;\in\; [0,1].
\end{equation}

The index is invariant to the choice of basis within each subspace, to sign flips, and to relabeling of modes within the block --- it measures the motion of the \emph{subspace}, not of any individual eigenvector, which is exactly the invariance required for a quantity that is to be interpreted as structural rather than as an artifact of eigenvector identification. Restricting to \(\mathcal{C}_t\) makes the measure immune to changes in index membership: entering and exiting names contribute no rotation.

Two conventions remain. The subspace dimension is three, justified spectrally in Section~\ref{sec:spectral} and by dose-response in Section~\ref{sec:dose}. The reported series is smoothed by a three-month moving average, a choice justified empirically in Section~\ref{sec:smoothing} rather than assumed here; the stored raw series is the object that the smoothing operates on, and the raw series alone already prices the premium (Section~\ref{sec:object}). Script {[}01{]} rebuilds the index from the panel and reproduces the stored series at correlation 1.0000, which is the gate condition for the remainder of the pipeline.

\subsection{What the index measures}\label{sec:measures}

Equation~\eqref{eq:rec} is a statement about geometry, and it has a reading in plain quantities that is exact rather than approximate. Because \(\cos^2\theta_i\) is the fraction of the \(i\)th direction that survives projection onto the new subspace, and the three squared cosines and squared sines partition unity direction by direction,

\begin{equation}\label{eq:retained}
1 - \mathrm{REC}_t \;=\; \tfrac{1}{3}\sum_{i=1}^{3}\cos^{2}\theta_i
\end{equation}

is \textbf{the fraction of the previous month's co-movement structure that the current month retains}. The index is therefore a proportion, not an abstract distance: a value of 0.194, the sample mean, says that a typical month rewrites about a fifth of the classification and carries four-fifths of it forward. Equivalently, since the three angles can be summarized by the single tilt whose squared sine equals the mean, the average month turns the subdominant frame by about 26 degrees.

The scale has a natural upper reference. Section~\ref{sec:calib} shows that two independently drawn three-dimensional subspaces of \(\mathbb{R}^{N}\) retain \(K/N \approx 0.007\) of each other, so a month that scrambled the structure entirely would register close to 1. Nothing in the sample comes near it. The largest single monthly reading is 0.526, in July 2002, which still carries just under half the previous structure forward.

\begin{figure}[htbp]
\centering
\includegraphics[width=0.98\textwidth]{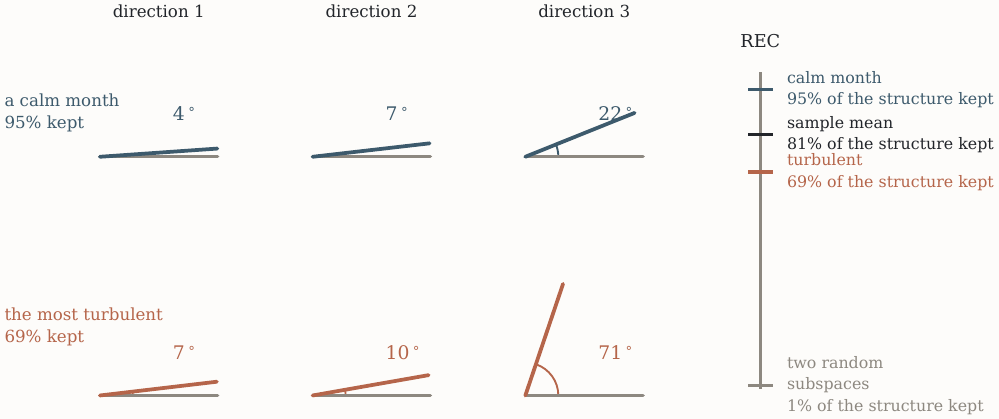}
\caption{What one observation of the index consists of [16]. \emph{Left:} the three principal angles between consecutive subdominant subspaces, for a quiet month (June 2016) and for the most turbulent month in the sample (March 2022). \emph{Right:} the index on its own scale, read as the share of the previous month's structure retained, with the random-subspace reference of Section~\ref{sec:calib} at the foot.}
\label{fig:measures}
\end{figure}

Figure~\ref{fig:measures} shows what a single observation consists of. The three principal angles are reported separately for a quiet month and for the most turbulent in the sample. In both, two of the three directions are nearly unmoved and the third carries most of the motion --- in March 2022 the first two turn by 7 and 10 degrees while the third turns by 71. The market does not rotate its classification uniformly; it holds most of the frame fixed and rewrites one axis at a time. Across the full sample the median angles are 5.9, 11.8 and 42.8 degrees.

This reading is used for exposition only. Every regression below takes the standardized index, so the scale and orientation of the raw series are immaterial to the estimates, and \(\mathrm{REC}\) rather than \(1-\mathrm{REC}\) is reported throughout because the mean squared sine is the standard projection metric on the Grassmannian and rises with the quantity being priced.

\subsection{Spectral placement}\label{sec:spectral}

Discarding modes five and higher requires a criterion. The usual one is the \citet{marchenko1967} edge for a null correlation matrix of aspect ratio \(q = T/N\), which for the median window (\(T = 12\), \(N = 429\)) sits at \((1+\sqrt{q})^2 = 1.362\) in the units of \eqref{eq:gram}. Against that edge, mode two is above the bulk in 67.9 percent of windows, mode three in 7.1 percent, and mode four never {[}09{]}. Taken literally this would recommend a one-dimensional or two-dimensional subspace and would make the phrase ``noise bulk dropped'' indefensible.

The raw edge is the wrong benchmark. The bulk of an empirical correlation matrix is not a unit-variance null: the market mode absorbs a large share of total variance and depresses the residual variance available to the remaining modes. Refitting the bulk after removing the leading mode, with

\begin{equation}\label{eq:edge}
\hat\sigma_t^{2} \;=\; \frac{\mathrm{tr}\,G_t - d_{1,t}}{T-1}, \qquad \text{edge}_t \;=\; \hat\sigma_t^{2}\,(1+\sqrt{q_t})^{2},
\end{equation}

gives a median residual variance of 0.750 and a median edge of 1.022 {[}09{]}. Table~\ref{tab:spectral} reports the resulting placement.

\begin{table}[htbp]
\centering
\caption{Spectral placement of the subdominant modes (368 windows) [09].}
\label{tab:spectral}
\small
\begin{tabular}{@{}crrr@{}}
\toprule
Mode & Median eigenvalue & Above raw MP edge (1.362) & Above deflated edge (median 1.022) \\
\midrule
2 & 1.544 & 67.9\% & 100.0\% \\
3 & 1.152 &  7.1\% &  97.3\% \\
4 & 0.967 &  0.0\% &  25.5\% \\
5 & 0.851 &  0.0\% &   0.8\% \\
6 & 0.771 &  0.0\% &   0.0\% \\
\bottomrule
\end{tabular}
\end{table}

Modes two and three are signal in essentially every window; mode four straddles the boundary, sitting above it in a quarter of windows; modes five and higher are bulk. The three-dimensional convention therefore contains two unambiguous modes and one marginal one, and the honest statement of the discard rule is against the deflated edge in \eqref{eq:retained}, never against the raw edge.

\subsection{Dose-response in the subspace dimension}\label{sec:dose}

The spectral classification makes a testable prediction about the pricing coupling. If modes two and three are signal, mode four is marginal and modes five onward are noise, then extending the subspace should improve the coupling up to dimension three and dilute it thereafter. Table~\ref{tab:modedim} rebuilds the index at subspace dimensions \(K = 2, \dots, 5\) and re-estimates the baseline regression of Section~\ref{sec:object} {[}10{]}.

\begin{table}[htbp]
\centering
\caption{Coupling by subspace dimension [10]. Dependent variable
$\mathrm{VRP}^{\log}$; regressors are the standardized three-month average of the
index, realized volatility and average correlation. Newey--West $t$, 12 lags, $n=365$.}
\label{tab:modedim}
\begin{tabular}{@{}ccrr@{}}
\toprule
Subspace dimension $K$ & Modes & $t$(REC) & $R^2$ \\
\midrule
2 & 2--3 & $+4.31$ & 0.479 \\
3 & 2--4 & $+5.40$ & 0.504 \\
4 & 2--5 & $+2.24$ & 0.442 \\
5 & 2--6 & $+1.97$ & 0.439 \\
\bottomrule
\end{tabular}
\end{table}

The profile is the predicted one. The two unambiguous modes carry the coupling on their own; adding the straddling mode strengthens it; adding the first clear bulk mode halves the statistic, and the second bulk mode weakens it further. One caveat is owed: part of the decline from \(K = 3\) to \(K = 5\) could reflect subspace-dimension mechanics, since the mean squared sine over a larger subspace is a different functional with a different null. Bulk dilution is the parsimonious reading and it is the one consistent with Table~\ref{tab:spectral}, but the two channels are not separated here.

\subsection{The Davis--Kahan channel, bounded}\label{sec:dk}

Eigenvector identification is governed by spectral gaps. The \citet{davis1970rotation} theorem, in the form convenient for statistics given by \citet{yu2015davis}, bounds the rotation of an estimated invariant subspace by the ratio of estimation error to the gap separating that subspace from the rest of the spectrum. When gaps compress, measured rotation increases mechanically even if nothing structural has moved. The index is therefore exposed to a specific and serious alternative: that it is a noisy transform of correlation intensity, since gaps compress when the market mode swells.

Script {[}06{]} tests the alternative directly. For each window we compute the pairwise-minimum gaps that bound the subdominant block --- \(d_1 - d_2\) above and \(d_4 - d_5\) below, each taken as the minimum across the two windows entering a rotation observation --- together with the average pairwise correlation and squared terms, and regress the raw index on all of them. The gaps enter with the negative signs Davis--Kahan predicts (\(t = -3.14\) for the upper gap, \(t = -2.15\) for the lower), confirming that the mechanical channel is real. It is nevertheless small: the full mechanical model explains \(R^2 = 0.121\) of the index's variance, leaving seven-eighths unaccounted for.

The decisive test is invariance rather than magnitude. Re-estimating the pricing regression of Section~\ref{sec:object} with the entire mechanical battery included as controls leaves the coupling numerically unchanged at \(t = +4.92\) {[}06{]}. Conditional sorts point the same way. Splitting the sample into quintiles of the lower gap, the coupling is \(+2.02\), \(+3.25\), \(+3.47\), \(+3.26\) and \(+0.97\) from the narrowest to the widest quintile: no monotone pattern, and the narrow-gap cell --- where the mechanical channel should be strongest --- is among the weaker ones. Sorting instead on the market mode's variance share, the coupling in the lowest quintile, that is in the calmest windows, is \(+3.96\), which rules out the reading that rotation is a shadow of stressed states. The coupling is not uniform across cells; it is absent in the middle quintile of the correlation sort (\(+0.77\)) and weak in the second and third quintiles of the variance-share sort (\(+1.67\), \(+1.76\)). What the sorts establish is the absence of the mechanical pattern, not homogeneity.

\subsection{Absolute calibration against a random-subspace null}\label{sec:calib}

Equation~\eqref{eq:rec} is bounded but has no natural scale: a mean squared sine of 0.19 is uninformative without a benchmark for what complete scrambling would look like. Two uniformly random three-dimensional subspaces of \(\mathbb{R}^{N}\) satisfy \(\mathbb{E}[\overline{\sin^{2}\theta}] = 1 - K/N\), which for \(K = 3\) and the average common cross-section \(N = 421\) equals 0.9929 {[}06b{]}; calibrating a geometric statistic against a random-subspace benchmark follows \citet{halperin2026a}. Retained alignment is the complement, \(\overline{\cos^{2}\theta}\), with a chance value of 0.0071.

Table~\ref{tab:kprofile} traces the \(k\)-horizon profile, comparing subspaces estimated \(k\) months apart.

\begin{table}[htbp]
\centering
\caption{$k$-horizon rotation against the random-subspace null [06b, 04].}
\label{tab:kprofile}
\begin{tabular}{@{}rrrrr@{}}
\toprule
$k$ (months) & $\overline{\sin^{2}\theta}$ & Retained alignment & Multiple of chance & $t$(VRP coupling) \\
\midrule
1  & 0.1944 & 0.8056 & $113.1\times$ & $+5.40$ \\
3  & 0.4307 & 0.5693 & $79.9\times$  & $+4.30$ \\
6  & 0.6578 & 0.3422 & $48.0\times$  & $+3.02$ \\
12 & 0.9234 & 0.0766 & $10.7\times$  & $+1.93$ \\
18 & 0.9274 & 0.0726 & $10.2\times$  & $+0.83$ \\
24 & 0.9336 & 0.0664 & $9.3\times$   & $+0.16$ \\
\bottomrule
\end{tabular}
\end{table}

\begin{figure}[htbp]
\centering
\includegraphics[width=0.98\textwidth]{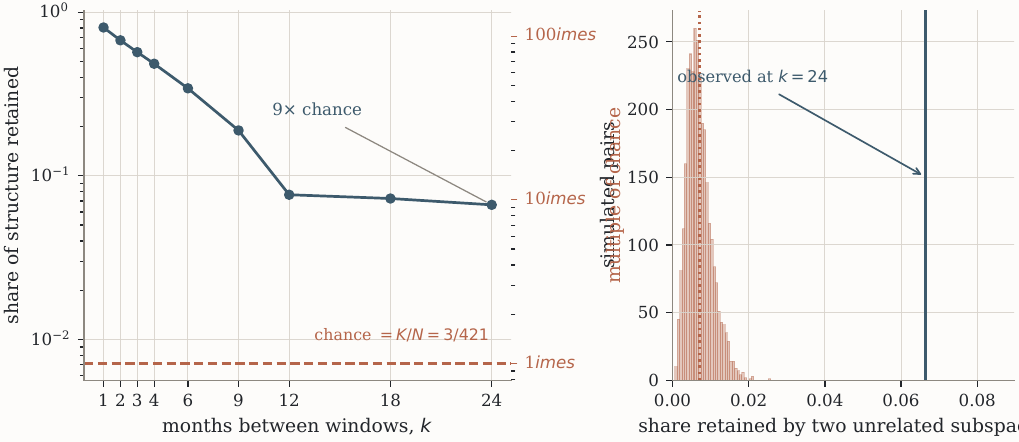}
\caption{\emph{Left:} the share of structure retained between subdominant eigenspaces estimated $k$ months apart, on a logarithmic scale, with the chance level as a dashed rule and a second axis reading the same curve in multiples of chance. Chance is $K/N=3/421$ because two independently drawn $K$-dimensional subspaces of $\mathbb{R}^{N}$ retain $K/N$ of each other in expectation. \emph{Right:} that statement verified by simulation. Three thousand pairs of independently drawn three-dimensional subspaces retain 0.0071 on average, exactly the analytic value, and the observed twenty-four-month figure lies far outside their entire range [16].}
\label{fig:kprofile}
\end{figure}

Two features matter. Rotation rises steeply out to roughly twelve months and then stops: the profile between \(k = 12\) and \(k = 24\) is nearly flat. And the plateau sits measurably below the null. A pair of subspaces separated by two years retains nine times the alignment that two random subspaces would share, and the multiple does not decay further over the horizons available. The decomposition this implies is a permanent structural core, on the order of seven percent subspace alignment and an order of magnitude above chance, plus a transient component with roughly twelve-month memory. The market neither settles onto a fixed classification of firms nor scrambles it; it perpetually revises the transient part while a core axis persists. Section~\ref{sec:discussion} identifies that core economically. Figure~\ref{fig:kprofile} plots the profile on a logarithmic scale with the chance level as a dashed rule.

%% file: sections/04_object.tex
\section{The object: coupling, orthogonality, spanning}\label{sec:object}

\subsection{The coupling}\label{sec:coupling}

The baseline specification regresses the premium on the standardized index with the two level measures as controls,

\begin{equation}\label{eq:baseline}
\mathrm{VRP}^{\log}_t \;=\; \alpha + \beta \,\widetilde{\mathrm{REC}}_t + \gamma_1 \,\mathrm{rv}_t + \gamma_2 \,\overline{\rho}_t + u_t,
\end{equation}

where \(\widetilde{\mathrm{REC}}\) denotes the standardized index and \(\overline{\rho}\) the average pairwise correlation. We say the index \emph{couples} to the premium when the regression association is significant and stable across specifications; the term is deliberately non-causal. Under the three-month convention, \(\hat\beta = +0.195\) with a Newey--West standard error of 0.036, \(t = +5.40\), a 95 percent confidence interval of \([0.124,\ 0.266]\), and \(R^2 = 0.504\) (\(n = 365\)); on the raw series, \(\hat\beta = +0.126\) with \(t = +4.02\) and \(R^2 = 0.456\) (\(n = 367\)) {[}02{]}. A one-standard-deviation increase in rotation is associated with a log premium higher by about a fifth, holding volatility and correlation levels fixed --- a fifth of a series whose own standard deviation is 0.697. The controls are not incidental: both enter negatively and significantly, so the coupling is identified against, not through, the level of volatility.

\subsection{Orthogonality}\label{sec:orthsec}

Table~\ref{tab:orth} reports the simple correlations between the index and every level measure available.

\begin{table}[htbp]
\centering
\caption{Correlations of the index with level gauges [02b].}
\label{tab:orth}
\begin{tabular}{@{}lrrr@{}}
\toprule
Gauge & Raw REC & 3-month REC & $n$ \\
\midrule
Realized volatility (12m equal-weight) & $-0.031$ & $+0.013$ & 365 \\
VIX                                    & $+0.180$ & $+0.316$ & 365 \\
Average pairwise correlation           & $-0.056$ & $-0.029$ & 365 \\
COR3M                                  & $+0.107$ & $+0.189$ & 240 \\
COR1M                                  & $+0.169$ & $+0.217$ & 240 \\
DSPX                                   & $+0.105$ & $+0.245$ & 139 \\
\bottomrule
\end{tabular}
\end{table}

The realized and average-correlation cells are effectively zero at both smoothings. The largest cell in the table is the smoothed index against VIX at 0.316, and it should be read as a warning rather than as a nuisance: roughly a tenth of shared variation with the implied-volatility level is precisely why every coupling regression in this paper controls for volatility and correlation levels, and why the Davis--Kahan battery of Section~\ref{sec:dose} is necessary rather than decorative. Smoothing raises every cell, which is consistent with the reading in Section~\ref{sec:wobble} that the monthly innovation carries a large re-estimation component.

\begin{figure}[htbp]
\centering
\includegraphics[width=0.6\textwidth]{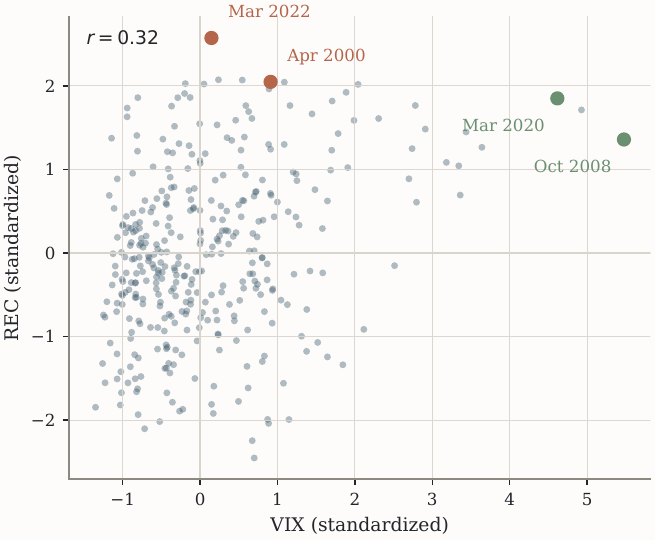}
\caption{The index against VIX, both standardized, 1995:06--2025:12 [02b, 16]. Four episodes are marked: two months of exceptional rotation at ordinary implied volatility, and two of exceptional implied volatility at moderate rotation.}
\label{fig:vsvix}
\end{figure}

Figure~\ref{fig:vsvix} shows the same relationship as a scatter, because the two episodes that make the distinctness concrete are invisible in a correlation. The highest rotation in the sample, March 2022, occurs at an entirely ordinary level of implied volatility; the most extreme implied volatility in the sample, October 2008, is accompanied by only moderate rotation.

\subsection{Spanning}\label{sec:spanning}

Correlation is a weak test of spanning. The stronger question is how much of the index's variation is recoverable from the traded implied-correlation surface. Table~\ref{tab:span} projects the index onto that surface, reporting the two available samples separately rather than conflating them.

\begin{table}[htbp]
\centering
\caption{$R^2$ from projecting the index on the implied-correlation surface [02b].}
\label{tab:span}
\begin{tabular}{@{}lrrrl@{}}
\toprule
Instruments & Raw REC & 3-month REC & $n$ & Sample \\
\midrule
COR3M, COR1M       & 0.059 & 0.050 & 240 & 2006:01--2025:12 \\
COR3M, COR1M, DSPX & 0.052 & 0.067 & 139 & 2014:06--2025:12 \\
\bottomrule
\end{tabular}
\end{table}

Between 5.0 and 6.7 percent of the index is spanned by the traded correlation surface, depending on smoothing and sample. Adding the dispersion index to the pair does not materially change the picture. Rotation is, to a first approximation, an unspanned state variable in the sense of \citet{collindufresne2002span}: it moves the premium without being recoverable from the instruments that price the premium's level. Section~\ref{sec:react} shows that this is not an artifact of instrument choice but follows from the algebra of what a level measure is, and pairs the spanning result with a reaction result that would otherwise appear to contradict it.

Eigenvector rotation is measured elsewhere, and the contribution claimed here is its pricing rather than its measurement. The distinction is also constructional. \citet{halperin2026b} reports a month-on-month rotation of the correlation matrix that correlates \(0.42\) with the irreversibility of a regime-aware ranking chain and near zero with its market-neutral counterpart --- that is, a rotation measure computed with the market mode retained, which therefore moves with market stress. The index used here strips the market mode and works in modes two to four, which is why it is orthogonal to the level measures of Table~\ref{tab:orth} where a market-inclusive rotation measure would not be, and why the validity battery of Section~\ref{sec:dk} is required rather than optional.

%% file: sections/05_priced.tex
\section{What is priced}\label{sec:priced}

\subsection{Frequency decomposition}\label{sec:freqsec}

Split each index into a persistent component, its three-month trailing average, and a high-frequency component, the raw series minus that average, and enter all four resulting series jointly with the level controls. Table~\ref{tab:freq} reports both premium conventions.

\begin{table}[htbp]
\centering
\caption{Frequency decomposition [02]. Newey--West $t$, 12 lags, $n=365$.
Controls: realized volatility, average correlation.}
\label{tab:freq}
\begin{tabular}{@{}lrr@{}}
\toprule
Component & $\mathrm{VRP}^{\log}$ & $\mathrm{VRP}^{\mathrm{lev}}$ \\
\midrule
Pearson, persistent      & $\mathbf{+4.92}$ & $\mathbf{+3.12}$ \\
Pearson, high-frequency  & $+0.30$ & $-0.28$ \\
Downside, persistent     & $+2.19$ & $+2.38$ \\
Downside, high-frequency & $+2.64$ & $+2.78$ \\
\midrule
$R^2$ & 0.526 & 0.192 \\
\bottomrule
\end{tabular}
\end{table}

The Pearson channel prices exclusively through its persistent part, in both conventions and at every specification we ran. The high-frequency component is not merely weaker; it is absent. The downside components appear priced here and are resolved in Section~\ref{sec:not}, where they do not survive mechanical controls.

\subsection{The smoothing plateau}\label{sec:smoothing}

A three-month average is a tuning choice, and a tuning choice that happens to maximize a test statistic is not evidence. Table~\ref{tab:smooth} sweeps the averaging window from one to twenty-four months in the baseline specification \eqref{eq:baseline}.

\begin{table}[htbp]
\centering
\caption{Smoothing sweep [02]. $t$(REC) in specification \eqref{eq:baseline}
as the averaging window $K$ varies.}
\label{tab:smooth}
\small
\begin{tabular}{@{}lrrrrrrrrrrr@{}}
\toprule
$K$ & 1 & 2 & 3 & 4 & 5 & 6 & 8 & 10 & 12 & 18 & 24 \\
\midrule
$t$   & 4.02 & 4.00 & \textbf{5.40} & \textbf{5.46} & \textbf{5.34} & \textbf{5.03} & 4.06 & 3.69 & 3.28 & 3.09 & 3.64 \\
$R^2$ & 0.456 & 0.473 & 0.504 & 0.508 & 0.510 & 0.505 & 0.485 & 0.474 & 0.465 & 0.464 & 0.466 \\
\bottomrule
\end{tabular}
\end{table}

\begin{figure}[htbp]
\centering
\includegraphics[width=0.72\textwidth]{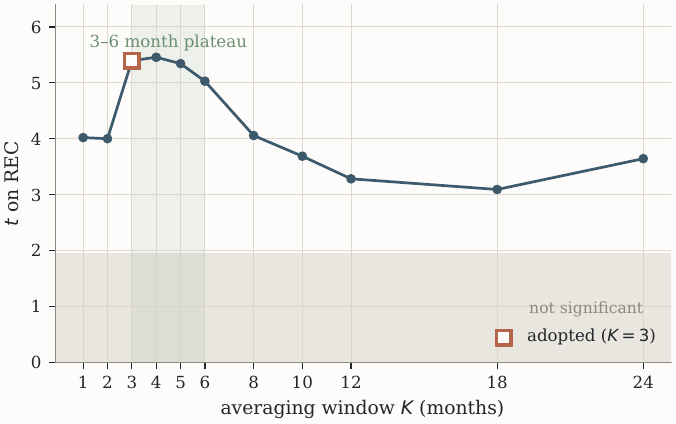}
\caption{The coupling across averaging windows [02, 16]. The shaded horizontal band marks values that would not reach conventional significance; no window falls into it. The vertical band marks the three-to-six-month plateau, and the open square marks the adopted convention, which sits at the plateau's left edge rather than at its maximum.}
\label{fig:smoothing}
\end{figure}

The coupling is significant at every window from one to twenty-four months, so no smoothing choice is load-bearing for the existence result. There is a flat plateau between three and six months, and the convention adopted sits at its left edge: \(K = 4\) and \(K = 5\) would both score marginally higher, which is the opposite of what a tuned choice looks like. Exponentially weighted averages with half-lives between two and six months are indistinguishable from the moving average (\(t \approx 5.4\)). Figure~\ref{fig:smoothing} plots the sweep.

\subsection{Why smoothing helps}\label{sec:wobble}

The plateau has a mechanical explanation that also sharpens the reading of Table~\ref{tab:freq}. Consecutive twelve-month windows share eleven months of data, so the one-month increment in the estimated subspace contains a substantial re-estimation component alongside whatever structural motion has occurred. The signature is visible in the autocorrelation function of the raw index: \(+0.035\) at lag one, \(-0.107\) at lag two, \(+0.181\) at lag three, \(+0.088\) at lag six and \(+0.363\) at lag twelve {[}02{]}. The negative second-order term is the reversal that wobble produces; the strong twelve-month term is the window length reappearing. A short moving average cancels wobble faster than it blurs genuine episodes, which is why the statistic rises from \(K=1\) to \(K=3\) and then decays as real episodes are averaged away.

The consequence for Table~\ref{tab:freq} is interpretive rather than statistical. The high-frequency component is contaminated by re-estimation noise, which is sufficient to account for its absence from the pricing regression. Whether it also carries structural content that simply goes unpriced is not settled here: the autocorrelation of the magnitude series is consistent with contamination but does not establish that the increment is noise-dominated, and separating the two would require a direct test on the direction of the increment rather than its size.

\subsection{Pace, not distance}\label{sec:pace}

Rotation can be accumulated over any horizon. Comparing subspaces \(k\) months apart yields a family of indices measuring how far the eigenspace has traveled over \(k\) months; Section~\ref{sec:calib} showed that this family saturates near \(k = 12\). The final column of Table~\ref{tab:kprofile} shows that the pricing coupling decays across the same family, from \(t = +5.40\) at one month to \(+1.93\) at twelve and \(+0.16\) at twenty-four.

Decay alone is ambiguous, since long-horizon rotation is a noisier and more saturated regressor. The discriminating test enters both ends of the family jointly. With the one-month and twelve-month indices in the same regression alongside the level controls, the short-horizon index retains \(t = +4.76\) while the twelve-month index falls to \(t = +1.38\) {[}04{]}. Short-horizon rotation drives out cumulative rotation, not the reverse.

Combined with the calibration of Section~\ref{sec:dk}, this identifies the priced object precisely. The permanent core is not compensated: it is a fixed feature of the market's structure and carries no premium. Neither is cumulative distance traveled. What is compensated is the \emph{rate} at which the transient component is being rewritten --- episodes of rapid revision, sustained for a quarter or two. The premium prices pace.

%% file: sections/06_mechanism.tex
\section{The mechanism: prepaid fear}\label{sec:mechanism}

\subsection{The two legs and one clock}\label{sec:legssec}

The premium in \eqref{eq:vrp} is a difference of two variances, and the coupling documented in Section~\ref{sec:object} must arise in one of them. Table~\ref{tab:legs} decomposes it, regressing each leg separately on the smoothed index and tracing the result forward.

One specification choice is forced. The realized leg of \eqref{eq:vrp} is \(\log(\mathrm{RV}_t/100)^2\), a monotone transform of the control \(\mathrm{rv}_t\), with which it correlates 0.9675 {[}03b{]}. Including realized volatility as a control in a regression whose dependent variable is realized variance regresses a variable on itself. The corrected specification therefore uses average correlation as the only control, and the uncontrolled regression is reported alongside it.

\begin{table}[htbp]
\centering
\caption{Leg decomposition and forward path [03b]. Dependent variables at $t+h$
regressed on the standardized three-month index at $t$ with average correlation as
control. Newey--West $t$, 12 lags, $n=365$ at $h=0$.}
\label{tab:legs}
\begin{tabular}{@{}rrrr@{}}
\toprule
$h$ & $t(\mathrm{VRP}^{\log})$ & $t(\log \mathrm{IV}^2)$ & $t(\log \mathrm{RV}^2)$ \\
\midrule
0  & $+5.12$ & $+4.57$ & $+0.62$ \\
1  & $+3.32$ & $+4.70$ & $+2.08$ \\
2  & $+2.08$ & $+4.54$ & $+2.75$ \\
3  & $+0.22$ & $+3.07$ & $+3.08$ \\
4  & $-1.11$ & $+1.92$ & $+3.23$ \\
5  & $-2.27$ & $+1.27$ & $+3.36$ \\
6  & $-2.54$ & $+1.36$ & $+3.64$ \\
7  & $-2.55$ & $+1.45$ & $+3.78$ \\
8  & $\mathbf{-2.68}$ & $+1.55$ & $\mathbf{+3.85}$ \\
9  & $-2.42$ & $+1.55$ & $+3.62$ \\
10 & $-1.38$ & $+1.77$ & $+2.96$ \\
11 & $-0.64$ & $+1.97$ & $+2.44$ \\
12 & $-0.41$ & $+1.69$ & $+2.10$ \\
\bottomrule
\end{tabular}
\end{table}

Without controls the contemporaneous split is \(+3.89\) on the implied leg and \(+0.07\) on the realized leg.

The pattern is a single clock running through both legs. Contemporaneously the entire association sits in implied variance; realized variance has not moved. Over the following quarters implied variance decays while realized variance builds, the two cross at about three months, and the premium --- their difference --- turns negative and troughs eight months out before dying at eleven. The premium widens when rotation is high and compresses once the volatility has arrived.

\subsection{The forward profile}\label{sec:profile}

The central claim requires a forward-looking test with proper controls for the persistence of volatility itself. Table~\ref{tab:horizon} regresses log realized variance \(h\) months ahead on the smoothed index with two of its own lags,

\begin{equation}\label{eq:forward}
\log \mathrm{RV}^2_{t+h} \;=\; \alpha_h + \beta_h \,\widetilde{\mathrm{REC}}_t + \phi_1 \log \mathrm{RV}^2_{t} + \phi_2 \log \mathrm{RV}^2_{t-1} + u_{t+h}.
\end{equation}

Specification \eqref{eq:forward} is the one that survives the \citet{valkanov2003} critique; the naive alternative that controls with volatility in levels leaves near-unit-root residuals and is documented as a failure in Appendix~\ref{app:inference}.

\begin{table}[htbp]
\centering
\caption{Forward realized-variance profile with AR(2) controls [03]. Bootstrap
$p$ from the simulated null of Section~\ref{sec:inference}, 1{,}000 replications.}
\label{tab:horizon}
\begin{tabular}{@{}rrrrr@{\hspace{2.4em}}rrrrr@{}}
\toprule
$h$ & $\hat\beta_h$ & $t$ & null 95\% & boot $p$ & $h$ & $\hat\beta_h$ & $t$ & null 95\% & boot $p$ \\
\midrule
1 & $+0.050$ & 4.09 & 2.52 & 0.003 & 7  & $+0.149$ & 3.81 & 2.34 & $<0.001$ \\
2 & $+0.079$ & 3.41 & 2.73 & 0.018 & \textbf{8} & $\mathbf{+0.155}$ & 3.65 & 2.29 & 0.001 \\
3 & $+0.101$ & 3.34 & 2.54 & 0.008 & 9  & $+0.149$ & 3.25 & 2.22 & 0.002 \\
4 & $+0.113$ & 3.41 & 2.50 & 0.007 & 10 & $+0.118$ & 2.42 & 2.16 & 0.025 \\
5 & $+0.125$ & 3.52 & 2.41 & 0.003 & 11 & $+0.096$ & 1.86 & 2.15 & 0.097 \\
6 & $+0.139$ & 3.76 & 2.40 & 0.002 & 12 & $+0.098$ & 1.82 & 2.13 & 0.098 \\
\bottomrule
\end{tabular}
\end{table}

The profile is a smooth hump. It clears the simulated null at every horizon from one to ten months, peaks at seven to eight months, and is dead at eleven and twelve. At the peak a one-standard-deviation increase in rotation is followed by log realized variance higher by 0.155, roughly seventeen percent in variance terms, two-thirds of a year later. Two features carry the interpretive weight. The effect at one month is significant but small, about a third of the peak, so the volatility consequences of reconfiguration are \emph{deferred} rather than immediate --- which is what makes the contemporaneous widening of the premium in Table~\ref{tab:legs} a prepayment rather than a contemporaneous repricing of visible risk. And the horizon at which predictability dies, eleven to twelve months, is the horizon at which Section~\ref{sec:calib} showed the subspace to have decorrelated. The forecast has exactly the memory of the object doing the forecasting.

The AR(4) robustness rerun executed for this draft (Appendix~\ref{app:inference}) leaves the coefficient path essentially unchanged, with a peak of \(+0.148\) at \(h = 8\), and clears the null at every horizon from one to nine (\(p \leq 0.027\)). At \(h = 10\) the bootstrap \(p\) rises to 0.077. The forecast claim is therefore made for one to nine months, where both lag specifications clear; the tenth month clears the primary design only.

\begin{figure}[htbp]
\centering
\includegraphics[width=0.74\textwidth]{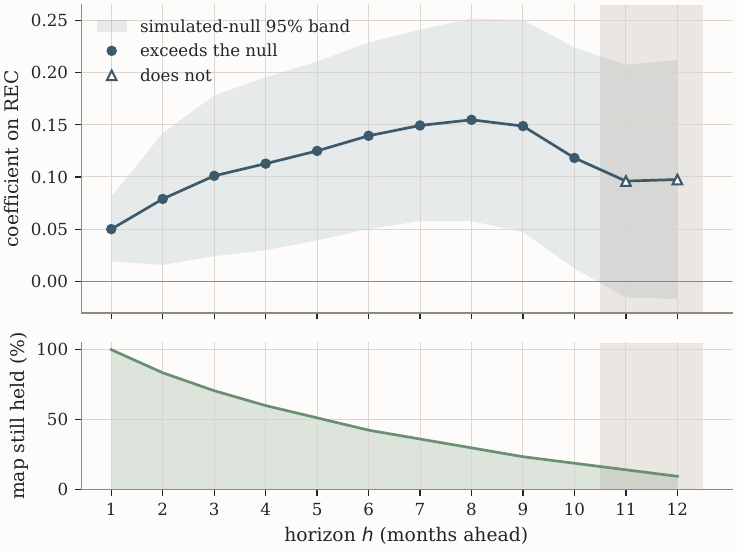}
\caption{\emph{Top:} coefficient path $\hat\beta_h$ from \eqref{eq:forward} with the simulated-null 95 percent band; filled circles mark horizons that exceed the null critical value, open triangles those that do not. \emph{Bottom:} the share of month-zero structure still held at each horizon, from the profile of Figure~\ref{fig:kprofile}. The forecast expires where the subspace has decorrelated, shaded at the right: predictive power crosses below the null at $h=11$, by which point roughly a tenth of the original map survives [16].}
\label{fig:hump}
\end{figure}

\subsection{Inference}\label{sec:inference}

Overlapping forward regressions on a persistent regressor invite two well-known distortions: the \citet{hansen1980forward} autocorrelation induced by overlap, addressed by Newey--West standard errors, and the \citet{stambaugh1999} bias induced by contemporaneous correlation between the regressor's innovation and the dependent variable's innovation. Newey--West alone is not enough here, and we do not rely on it.

The simulated null is constructed as follows. The dependent variable is generated from its own fitted AR(2), and the regressor is rebuilt by resampling raw-index innovations and passing them through the same three-month moving-average construction used on the real series, so that the null regressor has the estimator's autocorrelation structure by construction. The innovation \emph{pairs} are resampled jointly and independently across time. This preserves the contemporaneous correlation between the two innovations --- the Stambaugh channel, estimated at \(+0.066\) in the raw pairs --- while destroying every lead--lag relationship, which is exactly the null of no forward predictability. Empirical 95 percent critical values under this null range from 2.13 to 2.73 across horizons, above the nominal 1.96, confirming that Newey--West alone over-rejects mildly in this design. The null distributions are slightly positively centered, with means from \(+0.16\) to \(+0.85\) and the largest values at short horizons; the reported \(p\)-values are computed against the simulated distribution and therefore absorb that shift.

The analytic Stambaugh check agrees. The index's AR(1) coefficient is \(\hat\theta = 0.719\), the innovation correlation is \(+0.025\), and the implied bias is 0.2 percent of the estimated coefficient {[}03{]} --- three orders of magnitude below the effect. Appendix~\ref{app:inference} also documents a bootstrap design we rejected: resampling innovation pairs in \emph{blocks} preserves not only the contemporaneous correlation but also the lead--lag structure inside each block, which smuggles the alternative into the null and inflates the critical values until nothing is significant. The distinction is not cosmetic, and we report it because the block design is the more natural first choice.

\subsection{What is being forecast}\label{sec:forecast}

The scope condition of Section~\ref{sec:data} applies to the mechanism with full force. The forecast target in \eqref{eq:forward} is the twelve-month equal-weighted constituent volatility environment, not one-month capitalization-weighted index variance, against which the same index has \(|t| \leq 1.9\) at every horizon. The mechanism sentence in its exact form is therefore: structural flux raises option-implied fear immediately and forecasts a persistently elevated broad volatility environment over the subsequent two to three quarters; it does not forecast the short-horizon index variance-swap payoff. The second clause is not a weakness of the result but a condition of its consistency with the no-alpha boundary of Section~\ref{sec:alpha}. A state variable that forecast the variance-swap payoff would be a trading signal; this one forecasts the environment against which the premium is written.

%% file: sections/07_boundaries.tex
\section{What it is not}\label{sec:not}

Three pre-registered boundaries bound the claim, and each was specified before it was tested and is reported whether or not it flattered the index. A fourth subsection closes the section with a positive result that would otherwise appear to contradict Section~\ref{sec:spanning}, and a fifth records two further nulls established outside this paper.

\subsection{The tail channel is not separable at monthly resolution}\label{sec:tail}

If the variance premium is fundamentally a crash premium, the priced rotation should be the rotation of the \emph{downside} dependence structure rather than of the full-sample correlation structure. The test constructs a second index identically to \eqref{eq:gram}--\eqref{eq:rec} but from a both-down realized-semicovariance Gram in the sense of \citet{bollerslev2020semicov}: returns are replaced by their negative parts without demeaning, columns are normalized by downside semideviation, and the same dual-Gram, subspace and principal-angle pipeline is applied {[}01{]}. The result is a genuinely distinct object, correlating 0.224 with the Pearson index in raw form and 0.255 after smoothing.

The downside index prices the premium on its own, with \(t = +3.18\) smoothed and \(+3.88\) raw in specification \eqref{eq:baseline}. In a direct horse race, however, it does not dominate: with both indices and the level controls in the same regression, the Pearson index carries \(t = +4.83\) against the downside index's \(+2.23\). Under the levels premium with the unsmoothed index the ranking reverses; Appendix~\ref{app:robust}.3 records that cell and locates what drives it. The downside version is thus a weaker competitor from the outset, and its distinctive feature in Table~\ref{tab:freq} is not its persistent component but its high-frequency one, which is priced at \(+2.64\) in logs and \(+2.78\) in levels where the Pearson high-frequency component is dead.

That distinctive feature is the one that fails. Table~\ref{tab:downside} adds the mechanical determinants of downside activity --- the cross-sectional breadth of negative returns, realized semivariance, and the first differences of both --- to the four-component specification.

\begin{table}[htbp]
\centering
\caption{Collapse of the downside components under mechanical controls [05].
Dependent variable $\mathrm{VRP}^{\log}$, $n=365$.}
\label{tab:downside}
\begin{tabular}{@{}lrr@{}}
\toprule
Regressor & Baseline & $+$ mechanical controls \\
\midrule
Pearson, persistent          & $+4.92$ & $+4.16$ \\
Pearson, high-frequency      & $+0.30$ & $+0.52$ \\
Downside, persistent         & $+2.19$ & $+0.08$ \\
Downside, high-frequency     & $+2.64$ & $\mathbf{-2.31}$ \\
\midrule
Down-breadth                 & ---     & $+4.37$ \\
Realized semivariance        & ---     & $+5.17$ \\
$\Delta$ down-breadth        & ---     & $-1.05$ \\
$\Delta$ realized semivariance & ---   & $-4.59$ \\
\bottomrule
\end{tabular}
\end{table}

The downside persistent component goes to zero and the downside high-frequency component flips sign, while breadth and semivariance absorb the pricing. The Pearson persistent component passes through the same battery essentially intact. A sign flip of this kind is collinearity residue rather than economics: the downside high-frequency index and downside intensity were largely the same variable, and once intensity is conditioned on directly the residual is not interpretable.

A permutation placebo makes the mechanical channel visible. Shuffling returns cross-sectionally within each month preserves breadth, intensity and the volatility of every month exactly while destroying all persistent firm-level structure, so a downside index rebuilt on shuffled data measures nothing structural by construction. The comparison is run in a downside-only specification --- the two downside components with the level controls, without the Pearson components --- so the placebo and the actual index are scored identically; the resulting statistic is therefore not the \(+2.64\) of Table~\ref{tab:freq}, which comes from the four-component regression. Across 100 rebuilds, the structure-free index prices the premium at a mean \(t\) of \(+1.14\) with standard deviation 0.29 {[}05{]}. The actual index scores \(+1.97\) in that specification and exceeds the placebo distribution at \(p \approx 0.02\), so there is \emph{something} beyond the mechanical channel. Table~\ref{tab:downside} shows that the something is not separable from intensity once intensity is conditioned on directly.

\begin{figure}[htbp]
\centering
\includegraphics[width=0.7\textwidth]{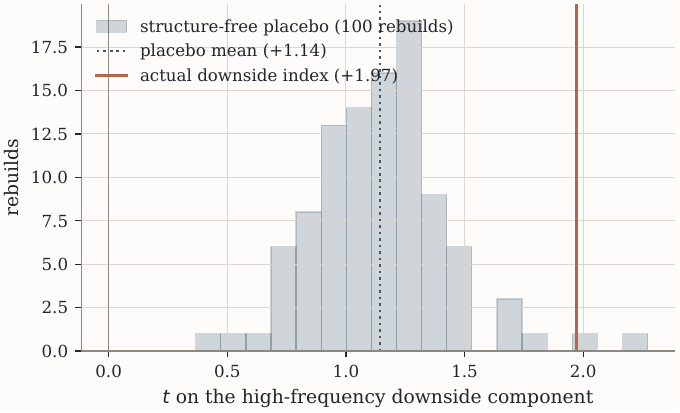}
\caption{Distribution of the high-frequency downside coefficient across 100 structure-free rebuilds, against the actual index [16].}
\label{fig:placebo}
\end{figure}

The verdict follows the pre-registered fork. With \(T = 12\) and roughly six down-observations per name per window, downside \emph{structure} and downside \emph{intensity} are not identified apart from each other: the semicovariance Gram is estimated from too few observations to distinguish which firms co-move on the downside from how many of them fell. The tail hypothesis is therefore not falsifiable at monthly resolution, and its apparent pricing is intensity. It is retired here rather than defended. Adjudication at daily resolution, where the observation count is not binding, is the companion paper's subject; the answer there is also negative, for a different and more informative reason having to do with which sampling frequency the premium selects.

\subsection{There is no timing alpha}\label{sec:alpha}

The index is a conditioning variable, not a signal. Scaling exposure to a short-variance position by the expanding-window standardized index, \(1 + 0.5 z_t\), and comparing the result to the unconditional position at equal volatility yields a paired \(t\) of \(-1.85\) across 329 months {[}08b{]} --- against the overlay, not for it. Unconditional and scaled Sharpe ratios are 1.49 and 1.36. Both figures are inflated by the non-tradeable payoff definition described in Section~\ref{sec:magnitude} and should be read only as a relative comparison. The conditioning information in the index does not translate into a timing rule. Section~\ref{sec:magnitude} shows that the same holds for the step and top-quartile forms the attribution invites; those are estimated on the shorter real-time-threshold sample of Appendix~\ref{app:harvest}.3, on which this linear rule scores \(-0.81\) rather than \(-1.85\), so the two figures are the same rule on different samples and not a discrepancy. The negative is not peculiar to this backtest: \citet{halperin2026b} reaches it independently with a real portfolio and realistic costs, finding that no correlation-concentration signal times market exposure better than the sign of the trailing market return out of sample, and that a contrarian re-risking rule which improved every window used to construct it failed on the one window held out of its construction.

\subsection{There is no crash protection}\label{sec:crash}

Nor is the index a hedge. Over the 329-month payoff sample, the worst five percent of months carry a mean standardized index reading of \(-0.47\) {[}08{]}: reconfiguration is \emph{below} average when short-variance positions suffer most. This is the P\&L-space form of the observation in Section~\ref{sec:data} that the index does not track the 2008--2009 episode proportionally. Whatever the premium is paying for, it is not insurance against the left tail of its own payoff. The boundary has a natural reading in the early-warning literature: correlation-geometry measures build up before crises that form endogenously inside the market and not before shocks that arrive from outside it \citep{quax2013}. \citet{halperin2026b} reports forward-drawdown areas under the ROC curve of 0.68 for 2008, 0.31 for 2020 and chance for 2001 on exactly such a measure. Rotation prices the ordinary variation of the premium, and the episodes it fails to anticipate are exactly the exogenous ones.

\subsection{The option market reacts without embedding}\label{sec:react}

Section~\ref{sec:spanning} showed that the traded implied-correlation surface spans at most 6.7 percent of the index. It does not follow that the option market is insensitive to reconfiguration. Regressing the slope of the implied-correlation curve, COR3M minus COR1M, on the smoothed index with the level controls gives \(t = -3.46\) and \(\hat\beta = -0.910\) index points per standard deviation, against a mean slope of \(+3.409\) across the 240-month sample {[}04{]}. The curve is normally upward-sloping, with far-dated implied correlation above near-dated; a one-standard-deviation increase in rotation removes about a quarter of that average slope, and the curve is inverted in 26.7 percent of the sample.

The two results are not in tension, and the precision matters because a careless statement of either would contradict the other. The \emph{level} of the correlation surface recovers 5 to 7 percent of the index's variation. The \emph{shape} of the same surface responds at \(t = -3.5\). A significant coupling and a small shared variance are perfectly compatible: the option market feels the state variable that its level instruments do not carry.

That the levels cannot carry it is not an accident of instrument choice but a property of what a level measure is. Any level functional of the correlation matrix is a Rayleigh quotient \(u^{\prime} C_t u\) for some weight vector \(u\), and decomposes as \(\sum_k \lambda_{k,t} (u^{\prime} v_{k,t})^2\). Because \(u\) is near-positive for every level measure in use --- equal weights on standardized returns for the average-correlation functional, volatility-scaled weights for index variance --- it overlaps the near-uniform market mode almost entirely and the sign-balanced body modes barely at all. Empirically, for the average-correlation functional the median \((u^{\prime}v_1)^2\) is 0.750 while the median body-mode overlaps are 0.013, 0.004 and 0.002, and the largest body overlap in any window is 0.432; for the index-variance functional the corresponding figures are 0.595 and a maximum body overlap of 0.353 {[}11{]}. The quotient is dominated by the \(\lambda_1\) term in the typical window, though not uniformly so, and the orientation of the body modes enters only through terms that are two orders of magnitude smaller. Level measures are therefore nearly blind to body rotation by construction. This argument explains the \emph{levels} spanning result only. The slope response is a different channel: it operates through expectations about the near term relative to the far term, which is not a contemporaneous functional of \(C_t\) at all.

\subsection{Two further nulls, not coded here}\label{sec:uncoded}

Two additional boundaries --- that the index does not improve covariance-matrix reliability out of sample, and that firm-level exposure to reconfiguration is not priced in the cross-section of returns --- are reported in the thesis from which this paper is drawn and are not coded in this repository. They are stated here as background rather than claimed as results of this paper, and their scripts are portable on request.

%% file: sections/08_discussion.tex
\section{Economics, discussion, limitations}\label{sec:discussion}

\subsection{Economic magnitude}\label{sec:magnitude}

The coupling is large in statistical terms; the question is whether it is large in economic ones. The natural accounting device is the variance harvest: at each month-end record the payoff to selling variance, \(\mathrm{IV}^2_t - \mathrm{RV}^2_{t+1}\), where \(\mathrm{IV}^2_t = (\mathrm{VIX}_t/100)^2\) and \(\mathrm{RV}^2\) is the paper's realized leg one month later. Two qualifications are required immediately. The payoff inherits the non-standard realized leg of Section~\ref{sec:data}, so it is not the payoff to a tradeable variance swap; and it ignores transaction costs, margin, and the convexity of the VIX-squared approximation. It is an attribution of the premium object this paper studies, not a strategy, and the numbers below should be read as decomposing where that object's compensation sits rather than as backtest results. Conditioning uses the expanding-window standardized index, with a 36-month burn-in and clipping at two standard deviations, so no observation uses information unavailable at the time. The sample runs from July 1998 to November 2025, 329 months, with an overall capture rate of 34.0 percent of implied variance sold {[}08{]}.

\begin{table}[htbp]
\centering
\caption{Variance-harvest attribution by index quartile [08, 08b]. $n=329$.
Q1 is the lowest and Q4 the highest quartile of the expanding-window
standardized index.}
\label{tab:harvest}
\footnotesize
\begin{tabular}{@{}lrrrrrr@{}}
\toprule
Quartile & Mean payoff & Capture & P\&L share & Worst month & $P(\text{loss})$ & Loss $\mid$ loss \\
\midrule
Q1 & $+0.0040$ & 11.1\% & 6\%  & $-0.099$ & 0.29 & $-0.0265$ \\
Q2 & $+0.0102$ & 27.4\% & 16\% & $-0.068$ & 0.21 & $-0.0164$ \\
Q3 & $+0.0092$ & 24.2\% & 14\% & $-0.064$ & 0.26 & $-0.0181$ \\
Q4 & $+0.0410$ & 52.7\% & \textbf{64\%} & $-0.032$ & 0.15 & $-0.0102$ \\
\bottomrule
\end{tabular}
\end{table}

Sixty-four percent of twenty-seven years of accumulated premium sits in the quarter of months with the highest rotation, and that quarter is simultaneously the safest: the loss frequency falls from 0.29 to 0.15, the mean loss conditional on a loss falls by three-fifths, and the worst single month improves from \(-0.099\) to \(-0.032\). Compensation and risk move in opposite directions across the sort, which is the signature of a conditioning variable rather than of a risk exposure.

The double sort sharpens the reading. Splitting first on the VIX median and then into rotation quartiles within each half, the rotation spread is flat when implied volatility is low (mean payoff \(+0.0032\) in Q1 against \(+0.0047\) in Q4) and wide when it is high (\(+0.0115\) against \(+0.0551\)) {[}08{]}. The sample's worst observation lives in the high-VIX, low-rotation cell. Rotation separates \emph{explained} fear, where implied volatility is elevated and a visible structural source accompanies it, from \emph{unexplained} fear, where implied volatility is elevated with no such source. Historically the first has been overpaid and safely collected; the second is where variance sellers are destroyed. This is attribution and conditioning. It is not alpha --- the timing overlay fails (Section~\ref{sec:alpha}) --- and it is not a hedge --- crash months carry below-average rotation (Section~\ref{sec:crash}).

The fourth-quartile concentration is the obvious invitation to a trading rule, and it does not convert into one under either functional form. A linear tilt, a step tilt keyed to the top quartile, and a top-quartile-only exposure all fail at equal volatility (\(t = -0.81\), \(-0.94\), \(-2.23\) on the expanding-threshold sample), and the step form evaluated on the second half of that sample is flat (\(-0.07\)) {[}08c{]}. The step was pre-committed as a diagnostic precisely because Table~\ref{tab:harvest} suggested it. The concentration is attribution, not extractable alpha, and Appendix~\ref{app:harvest} gives the full battery.

\begin{figure}[htbp]
\centering
\includegraphics[width=0.98\textwidth]{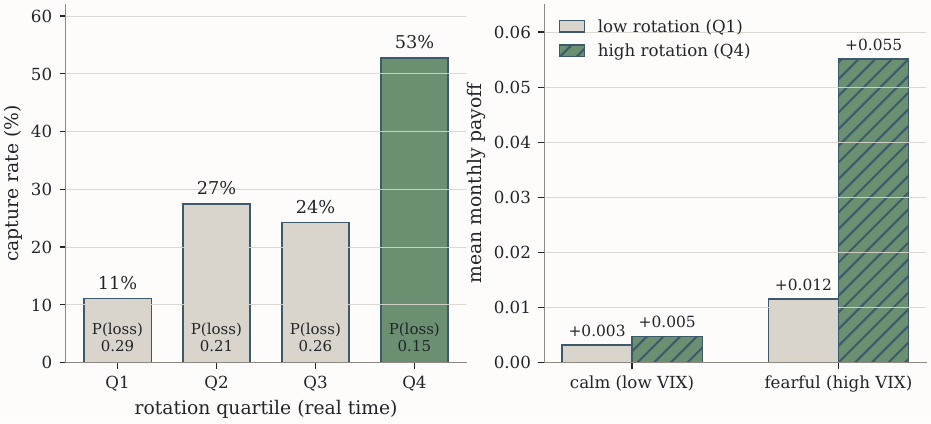}
\caption{Variance-harvest attribution [08, 16]. \emph{Left:} capture rate by real-time rotation quartile, with the loss frequency of each quartile printed inside the bar; compensation rises and risk falls together across the sort. \emph{Right:} the double sort. Quartiles are re-formed within each half of the VIX distribution. Rotation separates payoffs only when implied volatility is already elevated.}
\label{fig:economics}
\end{figure}

\subsection{The identity of the rotating structure}\label{sec:identity}

What is actually rotating? Labeling each subdominant mode in each window by the sector whose mean standardized loading is largest, subject to a magnitude threshold, gives a stable answer for the second mode: utilities in 63.6 percent of the 368 windows, energy in 17.7 percent, and no dominant sector in 13.0 percent {[}07{]}. The third mode is diffuse in 52.2 percent of windows and otherwise utilities or energy; the fourth is diffuse in 71.2 percent. The permanent core identified in Section~\ref{sec:calib} is therefore economically recognizable: it is the duration-and-defensives axis that separates rate-sensitive, low-beta names from the rest of the market, with a commodity axis appearing as its most frequent alternative. The same persistent utilities block is recovered independently by \citet{halperin2026a} from a different decomposition of the same market.

The identification of the core does not identify the priced motion, and three eliminations narrow it. First, the priced rotation is not discrete substitution of one labeled axis for another. Constructing a switch-intensity series from changes in the labeled sector composition of the three-mode block gives a correlation with the raw index of \(+0.187\), and entering the two jointly with the level controls leaves the index at \(t = +4.89\) against the switch series at \(+1.93\) {[}07{]}. Label churn is correlated with rotation and does not absorb it. Second, the priced rotation is not spectral-gap compression: Section~\ref{sec:dk} bounds that channel at 12.1 percent of the index's variance with the coupling invariant. Third, the cross-sectional churn of the volatility hierarchy, measured as mean absolute daily change in volatility percentile rank following the volatility-ranking construction of \citet{halperin2026a}, is orthogonal to the index and unpriced, leaving the coupling intact {[}07b{]}. This is the one test in the paper estimated on the daily Bloomberg panel rather than the monthly one; because that panel is licensed, this result alone is not reproducible from the public repository.

The eliminations are informative because they map onto the decomposition of Section~\ref{sec:intro}. Writing \(C_t\) as generated by loadings and factor covariance, the risk-mix channel --- rotation caused by changing relative factor volatilities with loadings fixed --- is what volatility-rank churn measures, and it is not the priced object. Discrete axis substitution is not the priced object either. What remains inside \(\dot V_t\) is \(\dot B_t\): continuous, diffuse revision of which firms load on the operative classification axes. Figures 9 and 10 show what that classification looks like and how far it moves. In April 2000 the leading subdominant mode separates technology from utilities, the growth-against-defensive split of that period; by March 2022 it separates energy from technology, and utilities has been displaced to the third mode. The same firms and the same arithmetic yield a different organizing question. Across the full sample the identifiability of the modes decays down the block, from a second mode that carries a dominant sector in 87 percent of windows to a fourth that is diffuse in 71 percent --- an independent corroboration of the mode-count evidence in Section~\ref{sec:dose}.

\begin{figure}[htbp]
\centering
\includegraphics[width=0.98\textwidth]{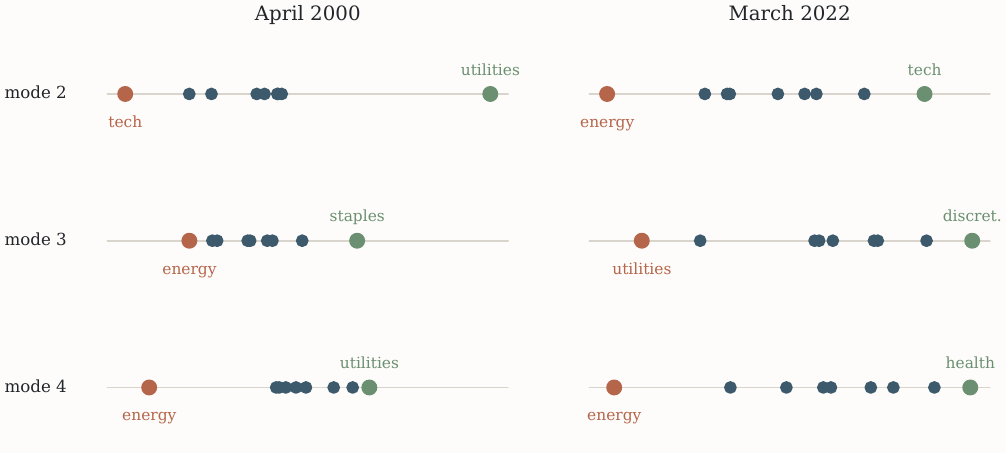}
\caption{The three subdominant modes as sector spectra, at two dates [07, 16]. Each line places the nine sector means on one mode; the extreme sectors are marked. Signs are fixed by a rule and carry no meaning beyond which sectors lie at opposite ends.}
\label{fig:axes}
\end{figure}

\begin{figure}[htbp]
\centering
\includegraphics[width=0.76\textwidth]{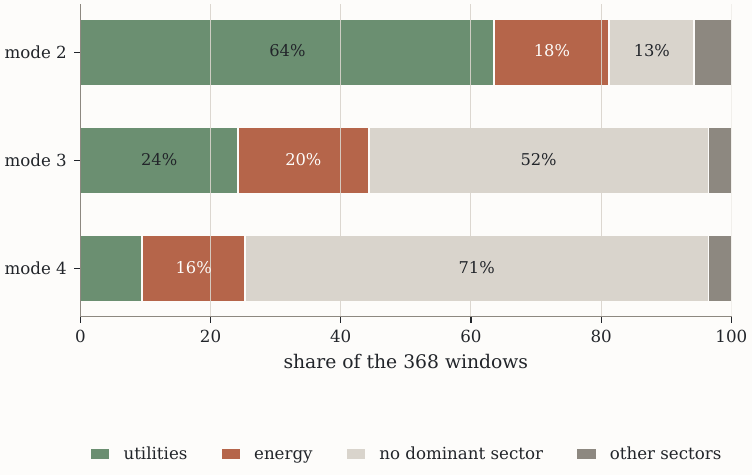}
\caption{Sector composition of the subdominant modes across the 368 windows [07, 16]. A window is assigned to the sector whose mean standardized loading on that mode is largest, subject to a magnitude threshold, and to ``no dominant sector'' otherwise.}
\label{fig:sectors}
\end{figure}

By elimination within this decomposition, then, the priced object is the rate at which the market rewrites its classification of firms. The identification is negative rather than direct: it rests on the two channels ruled out above rather than on a measurement of loading drift itself, and Section~\ref{sec:discussion}.4 records that limitation.

This object has a name in practitioner language. The market-removed residual of the correlation matrix is what the crowding literature measures \citep{baltas2019,zlotnikov2024}, and drift in the loadings on the operative axes is drift in which names are crowded together. Read that way, \(\dot B_t\) is the turnover of crowding structure, and the premium documented here is the compensation for holding variance exposure while that structure is being rewritten.

\subsection{Frequency positioning}\label{sec:freqpos}

The layer of geometry studied here is monthly-sampled, and that is not innocuous. Daily-sampled subspace geometry is a distinct object with different statistical properties, and the companion paper shows that it is not priced. The relevant lineage runs from \citet{epps1979} and \citet{bonanno2001hf}, who established that measured correlation structure depends systematically on sampling frequency, through the heterogeneous-market tradition of \citet{muller1997volatilities} and the cascade models of \citet{corsi2009har}, to the frequency-domain asset pricing of \citet{dewbecker2016frequency}. If agents operating at different horizons write structure at their own horizons, then correlation geometry is layered by frequency, and a premium may select one layer. The result here and in the companion paper is that the premium selects the layer written at the horizon on which institutional portfolios are actually reallocated, and is indifferent to the layer written by intraday and daily trading. That is a claim about which frequency carries economic content, and it is developed properly in the companion paper.

\subsection{Limitations}\label{sec:limits}

The premium in \eqref{eq:vrp} has mismatched legs in horizon and weighting; this is disclosed in Section~\ref{sec:data} as a scope condition and every result is a statement about that object rather than about the canonical variance risk premium. The analysis universe is survivor-tilted by the coverage filter (Appendix~\ref{app:panel}), so the realized leg describes the persistent large-capitalization cross-section rather than the index as constituted at each date. Evidence comes from a single market and a single premium. The implied-correlation results rest on the shorter post-2006 and post-2014 subsamples, where 240 and 139 monthly observations limit what can be asked. The downside verdict of Section~\ref{sec:tail} is a statement about an estimator at \(T = 12\), not about the world: it asserts that monthly data cannot separate downside structure from downside intensity, not that the tail hypothesis is false. Simulated-null inference is conducted under AR(2) with an AR(4) robustness rerun; the tenth-month horizon is significant under the former and marginal under the latter. And the identification of \(\dot B_t\) in Section~\ref{sec:identity} proceeds by elimination within a decomposition rather than by direct measurement of loading drift, which would be the natural next construction. Finally, the negative results of Section~\ref{sec:not} are consistent with a broader pattern: the early-warning literature on correlation and network structure has repeatedly found measures that describe the cross-section well contemporaneously and time it poorly out of sample \citep{george2023}, and nothing here should be read as an exception to that.

\subsection{Closing}\label{sec:closing}

Two facts about the market's co-movement structure sit uneasily together until they are measured on the same scale. The structure never settles: the transient component of the subdominant eigenspace turns over on a roughly twelve-month clock, and the market's classification of firms at any date bears only limited resemblance to its classification a year earlier. And the structure is never forgotten: a permanent core, economically the duration-and-defensives axis, holds at roughly ten times chance alignment indefinitely. The market neither learns its co-movement structure nor forgets it. It perpetually revises --- and the variance risk premium is compensation for the pace of revision (\(t \approx 5\)), paid ahead of the volatility that the revision brings.

%% file: appendices/A_inference.tex
\section{Inference designs}\label{app:inference}

\textbf{A.1 The naive-controls failure.} The natural first specification for \eqref{eq:forward} controls with contemporaneous volatility and correlation levels rather than with lags of the dependent variable. It is misspecified. At \(h = 1\), where there is no overlap and residual autocorrelation must therefore come from the specification itself, the naive regression leaves a residual first-order autocorrelation of \(+0.502\); the AR(2) specification leaves \(+0.002\) {[}14{]}. At longer horizons both designs show high residual autocorrelation, but there the source is overlap and Newey--West is the appropriate correction. The naive design is the configuration \citet{valkanov2003} identifies as unreliable: a persistent regressor, a persistent dependent variable, and residuals that inherit a near-unit root. Its coefficients are similar to the reported ones (\(t = +2.53\), \(+3.03\), \(+3.57\), \(+3.83\), \(+2.10\) at \(h = 1, 3, 6, 8, 12\)), but they are not trustworthy and are not the basis of any claim in this paper.

\textbf{A.2 The i.i.d.-pairs null.} The reported bootstrap generates the dependent variable from its fitted AR(2) and rebuilds the regressor by resampling raw-index innovations and passing them through the three-month moving-average construction. Innovation pairs \((v_t, w_t)\) are drawn jointly and independently across time, which preserves their contemporaneous correlation --- the Stambaugh channel --- while destroying all lead--lag structure. One thousand replications give 95 percent critical values of 2.13 to 2.73 and 99 percent values of 2.84 to 3.66 across horizons {[}03{]}.

\textbf{A.3 The block-pairs pitfall.} Resampling the same pairs in blocks, as a stationary or moving-block bootstrap would \citep{politis1994stationary}, is the more natural first choice and is wrong here. Blocks preserve the within-block lead--lag structure, which is precisely the alternative hypothesis; the resulting null distribution contains the effect being tested. With blocks of twelve months the 95 percent critical values rise to between 3.63 and 5.59, against 2.12 to 2.47 for the i.i.d.-pairs design, and no horizon in Table~\ref{tab:horizon} would be declared significant {[}14{]}. We report the failed design because the diagnostic --- a null whose critical values are several times the nominal value --- is the signature of a bootstrap that has absorbed its own alternative.

\textbf{A.4 Stambaugh algebra.} With \(\hat\theta = 0.719\) for the smoothed index's AR(1) coefficient and an innovation correlation of \(+0.025\), the standard bias approximation \(-\left(\sigma_{uv}/\sigma_v^2\right)\left(1+3\hat\theta\right)/n\) evaluates to 0.2 percent of the estimated coefficient at \(n = 365\) {[}03{]}. The bias is negligible relative to the effect, and the simulated null preserves the channel in any case.

\textbf{A.5 AR(4) robustness.} Table~\ref{tab:ar4} reruns Table~\ref{tab:horizon} with four lags of the dependent variable and the correspondingly re-specified null {[}12{]}.

\begin{table}[htbp]
\centering
\caption{Forward profile under AR(4) controls, 1{,}000 replications [12].}
\label{tab:ar4}
\begin{tabular}{@{}rrrrr@{\hspace{2.4em}}rrrrr@{}}
\toprule
$h$ & $\hat\beta_h$ & $t$ & null 95\% & boot $p$ & $h$ & $\hat\beta_h$ & $t$ & null 95\% & boot $p$ \\
\midrule
1 & $+0.049$ & 3.96 & 2.09 & $<0.001$ & 7  & $+0.144$ & 3.43 & 2.57 & 0.003 \\
2 & $+0.079$ & 3.28 & 2.39 & 0.009 & 8  & $+0.148$ & 3.25 & 2.43 & 0.009 \\
3 & $+0.100$ & 3.16 & 2.65 & 0.027 & 9  & $+0.141$ & 2.86 & 2.35 & 0.017 \\
4 & $+0.110$ & 3.18 & 2.80 & 0.030 & 10 & $+0.109$ & 2.12 & 2.37 & 0.077 \\
5 & $+0.122$ & 3.25 & 2.72 & 0.022 & 11 & $+0.093$ & 1.75 & 2.39 & 0.148 \\
6 & $+0.136$ & 3.44 & 2.67 & 0.006 & 12 & $+0.095$ & 1.71 & 2.38 & 0.147 \\
\bottomrule
\end{tabular}
\end{table}

The coefficient path is essentially unchanged, peaking at \(+0.148\) at \(h = 8\) against \(+0.155\) under AR(2). Horizons one through nine clear the null at \(p \leq 0.027\); the fourth-horizon value of 0.030 sits at the boundary. The tenth horizon does not clear (\(p = 0.077\)), which is the single substantive difference between the two designs and is reflected in the claim made in Section~\ref{sec:profile}.

%% file: appendices/B_robustness.tex
\section{Robustness}\label{app:robust}

\textbf{B.1 The levels premium.} Every headline result is reported in both conventions in the main text; the levels premium is weaker throughout but never sign-reversed. In the baseline specification \eqref{eq:baseline}, the levels coupling is \(t = +3.21\) with \(R^2 = 0.149\) under the three-month convention and \(t = +2.62\) with \(R^2 = 0.078\) on the raw index, against \(+5.40\) and \(+4.02\) in logs. The smoothing sweep in levels gives \(t = +2.62\), \(+3.21\), \(+3.23\), \(+3.40\), \(+1.61\) at \(K = 1, 3, 4, 6, 12\): the same plateau shape, shifted right and attenuated. The gap between conventions is the 2008--2009 domination documented in Section~\ref{sec:data} and is treated there as evidence about what the index prices rather than as a specification problem.

\textbf{B.2 Exponential smoothing.} Replacing the moving average by an exponentially weighted average leaves the result unchanged: \(t = +4.91\), \(+5.42\), \(+5.47\), \(+5.44\), \(+5.38\), \(+5.33\), \(+5.28\) at half-lives of one, two, three, four, six, nine and twelve months. Half-lives between two and six months are indistinguishable at \(t \approx 5.4\), matching the moving-average plateau of Table~\ref{tab:smooth} and confirming that the persistence result is not an artifact of the averaging kernel.

\textbf{B.3 The downside index under the levels convention.} The setup of Section~\ref{sec:tail} is convention-dependent in a way worth recording. Under the levels premium with the raw index, the downside index prices at \(t = +2.98\) on its own and wins the horse race against the Pearson index, \(+2.78\) against \(+2.36\). Under the levels premium with the smoothed index the ranking has already reversed (\(+2.39\) against \(+3.12\)), and under the paper's primary log convention it reverses decisively (\(+2.23\) against \(+4.83\)). The downside index's apparent advantage is thus confined to the levels-and-unsmoothed cell --- the cell in which, by Section~\ref{sec:data}, the 2008--2009 episode dominates the dependent variable, and in which the tail reading of the premium is therefore most flattered by construction.

The verdict of Section~\ref{sec:tail} does not depend on the convention. Rerunning the full mechanical battery of Table~\ref{tab:downside} against the levels premium reproduces the collapse: the downside persistent component falls to \(t = +0.21\) and the downside high-frequency component flips to \(-2.21\), while realized semivariance absorbs the pricing at \(+6.72\) and the Pearson persistent component survives at \(+2.80\). The two conventions bracket the range quoted in Section~\ref{sec:tail} for the downside persistent component, 0.08 in logs and 0.21 in levels.

%% file: appendices/C_harvest.tex
\section{The variance-harvest attribution}\label{app:harvest}

\textbf{C.1 Construction.} The payoff recorded at each month-end is \(\mathrm{IV}^2_t - \mathrm{RV}^2_{t+1}\) with \(\mathrm{IV}^2_t = (\mathrm{VIX}_t/100)^2\) and \(\mathrm{RV}^2_{t+1}\) the paper's realized leg one month forward. Conditioning uses the expanding standardized index with a 36-month burn-in, clipped at two standard deviations. The sample is July 1998 to November 2025, 329 months.

\textbf{C.2 Double sort.} Splitting on the VIX median and then into rotation quartiles within each half gives mean payoffs of \(+0.0032\) (Q1) and \(+0.0047\) (Q4) in the low-VIX half, with worst months of \(-0.031\) and \(-0.024\); and \(+0.0115\) (Q1) and \(+0.0551\) (Q4) in the high-VIX half, with worst months of \(-0.099\) and \(-0.024\) {[}08{]}. The rotation sort is informative only where implied volatility is already elevated.

\textbf{C.3 The timing battery.} Table~\ref{tab:timing} tests whether the quartile concentration of Table~\ref{tab:harvest} converts into a rule. Quartile membership is assigned in real time from the expanding-window 75th percentile of the standardized index computed on history strictly prior to each month, which costs a further 36-month burn-in and leaves 293 of the 329 payoff months; 69 of them (24 percent) are flagged top-quartile in real time. Each rule is compared to the unconditional position after rescaling to equal volatility, and the reported statistic is the paired \(t\) on the difference.

\begin{table}[htbp]
\centering
\caption{Timing rules against the unconditional short-variance position [08c]. $n=293$.}
\label{tab:timing}
\begin{tabular}{@{}llrr@{}}
\toprule
Rule & Weight & Sharpe & Equal-volatility $t$ \\
\midrule
Unconditional      & $1$                                  & 1.28 & --- \\
Linear tilt        & $1 + 0.5\,z_t$                       & 1.22 & $-0.81$ \\
Step tilt          & $1 + 0.5\cdot\mathbf{1}\{Q4\}$       & 1.25 & $-0.94$ \\
Top quartile only  & $\mathbf{1}\{Q4\}$                   & 1.02 & $-2.23$ \\
\bottomrule
\end{tabular}
\end{table}

Split in half, the step form gives \(-0.97\) on 2001:07--2013:08 and \(-0.07\) on 2013:09--2025:11. The step was pre-committed as a diagnostic because Table~\ref{tab:harvest} suggested it, and it is reported whatever it shows. None of the three forms improves on the unconditional position, and the third --- which concentrates the book into the top quartile --- is significantly worse.

\textbf{C.4 Caveats.} Three qualifications limit any strategic reading. The realized leg is the paper's non-standard object, so the payoff is not that of a tradeable variance swap; a genuine one-month variance-swap payoff would use the canonical construction against which the index is not priced (Section~\ref{sec:forecast}). Selling variance at \(\mathrm{VIX}^2\) ignores the convexity correction relating the index to the fair variance-swap strike \citep{demeterfi1999,brittenjones2000}, as well as transaction costs and margin. And the Sharpe ratios quoted in Section~\ref{sec:alpha} inherit both distortions. The tables are an attribution of where the premium studied in this paper has historically been earned, and nothing more.

%% file: appendices/D_panel.tex
\section{Composition of the estimation panel}\label{app:panel}

\textbf{D.1 Coverage.} The panel carries 159,742 monthly return observations, 98.0 percent of the \(379 \times 430\) rectangle, with a pooled mean of \(+1.24\) percent, a standard deviation of 10.82 percent, and first and ninety-ninth percentiles of \(-25.5\) and \(+31.2\) percent {[}15{]}.

\begin{table}[htbp]
\centering
\caption{Panel coverage. Entries are names whose first observation falls in the period;
exits are names whose last observation falls in it [15].}
\label{tab:coverage}
\small
\begin{tabular}{@{}lrrrrrr@{}}
\toprule
Period & Months & Mean names & Min & Max & Entries & Exits \\
\midrule
1994 (from June) &   7 & 348.4 & 344 & 358 & 14 & 0 \\
1995             &  12 & 364.9 & 358 & 376 & 18 & 0 \\
1996             &  12 & 380.4 & 378 & 384 &  8 & 0 \\
1997             &  12 & 390.1 & 385 & 398 & 14 & 0 \\
1998             &  12 & 405.7 & 399 & 410 & 12 & 0 \\
1999             &  12 & 412.8 & 410 & 419 & 10 & 0 \\
2000             &  12 & 424.8 & 419 & 429 & 10 & 0 \\
2001--2017       & 204 & 429.6 & 429 & 430 &  0 & 0 \\
2018             &  12 & 429.5 & 429 & 430 &  0 & 0 \\
2019             &  12 & 429.2 & 429 & 430 &  0 & 1 \\
2020--2022       &  36 & 429.0 & 429 & 429 &  0 & 0 \\
2023             &  12 & 428.8 & 428 & 429 &  0 & 1 \\
2024             &  12 & 426.9 & 426 & 428 &  0 & 0 \\
2025             &  12 & 424.6 & 422 & 427 &  0 & 6 \\
\bottomrule
\end{tabular}
\end{table}

\textbf{D.2 The survivorship disclosure.} The retained universe is 430 names, of which 344 are present in the first month and 86 enter later, all before 2001. Eight names end before the final month: two mid-sample, in 2019:02 and 2023:10, and six during 2025, the latter being sample-edge turnover rather than economic exit. Median per-name coverage is 100 percent and the minimum is 74.7 percent.

The disclosure this requires is plain. The source panel includes delisted names, but the 70/80 coverage filter of Section~\ref{sec:data} retains only names with long histories, and the result is a near-constant-membership panel with almost no in-sample attrition. The analysis universe is therefore survivor-tilted by construction. Two consequences follow. The equal-weighted realized leg is the volatility environment of the persistent large-capitalization cross-section rather than of the index as constituted at each date, which is a further respect in which the premium of Section~\ref{sec:data} is not the canonical object. And the rotation measure is estimated on a cross-section that does not itself churn, so what it records is drift in the co-movement of a broadly fixed set of firms rather than turnover in index membership --- the intended object, but a property of the filter as much as of the market. The full constituent list is supplied rather than summarized: \texttt{ticker\_inventory.csv} records all 430 retained names with the first and last month each is observed and its coverage share. It is distributed as an ancillary file with this preprint and in the replication repository, so the exact universe underlying every estimate can be reconstructed without re-deriving the filter. Printing the list here would consume several pages to no purpose that the file does not serve better.

\textbf{D.3 Sensitivity to the coverage thresholds.} Table~\ref{tab:filter} rebuilds the index under alternative column and row filters and re-estimates the baseline coupling of Section~\ref{sec:object}.1 {[}17{]}.

\begin{table}[htbp]
\centering
\caption{Panel filter sensitivity [17]. Column threshold is the minimum share of
months a name must be observed; row threshold the minimum share of names a month
must carry. The final column is the correlation of the raw monthly index with the
baseline index over common months.}
\label{tab:filter}
\small
\begin{tabular}{@{}lrrrrrr@{}}
\toprule
Column / row & Months & Names & $n$ & $t$(REC) & $R^2$ & Corr.\ baseline \\
\midrule
0.60 / 0.90 & 314 & 464 & 300 & $+6.11$ & 0.504 & 0.945 \\
0.60 / 0.80 & 362 & 464 & 348 & $+5.68$ & 0.495 & 0.952 \\
0.70 / 0.70 & 409 & 431 & 368 & $+5.24$ & 0.499 & 0.998 \\
\textbf{0.70 / 0.80} (adopted) & 379 & 430 & 365 & $+5.40$ & 0.504 & 1.000 \\
0.70 / 0.90 & 346 & 430 & 332 & $+5.62$ & 0.485 & 1.000 \\
0.80 / 0.80 & 403 & 385 & 368 & $+4.46$ & 0.468 & 0.917 \\
0.90 / 0.80 & 430 & 326 & 368 & $+2.44$ & 0.443 & 0.833 \\
\bottomrule
\end{tabular}
\end{table}

The coupling is significant at every pair, from \(+2.44\) to \(+6.11\), and the adopted pair is not the maximum --- a looser column filter with a stricter row filter scores higher. The one configuration that materially weakens the result is the tightest column threshold, which cuts the cross-section to 326 names and correlates only 0.833 with the baseline index; that is the breadth cost the rationale in Section~\ref{sec:data} anticipates, and it points in the opposite direction from a tuned choice.

%% file: sections/09_prior_work.tex
\section{Note on prior work}

The authors' prior publication in this lineage is \citet{carvalho2024etf}, which applies stochastic-geometry techniques and minimum-spanning-tree analysis to a panel of 85 exchange-traded funds. It contains no reconfiguration index, no rotation measure, no variance-risk-premium results and no pre-registered nulls. It is cited here, together with \citet{vilelamendes2002process}, \citet{vilelamendes2003reconstructing} and \citet{araujo2007geometry}, as method lineage only --- for the correlation-geometry framework and the surrogate-based approach to eigenvalue significance. Every claim about the reconfiguration index is established inside this paper.

%% file: main.bbl
\begin{thebibliography}{40}
\providecommand{\natexlab}[1]{#1}
\providecommand{\url}[1]{\texttt{#1}}
\expandafter\ifx\csname urlstyle\endcsname\relax
  \providecommand{\doi}[1]{doi: #1}\else
  \providecommand{\doi}{doi: \begingroup \urlstyle{rm}\Url}\fi

\bibitem[Ara{\'u}jo and Lou{\c{c}}{\~a}(2007)]{araujo2007geometry}
Tanya Ara{\'u}jo and Francisco Lou{\c{c}}{\~a}.
\newblock The geometry of crashes: A measure of the dynamics of stock market
  crises.
\newblock \emph{Quantitative Finance}, 7\penalty0 (1):\penalty0 63--74, 2007.

\bibitem[Baltas(2019)]{baltas2019}
Nick Baltas.
\newblock The impact of crowding in alternative risk premia investing.
\newblock \emph{Financial Analysts Journal}, 75\penalty0 (3):\penalty0 89--104,
  2019.

\bibitem[Bj{\"o}rck and Golub(1973)]{bjorck1973angles}
{\AA}ke Bj{\"o}rck and Gene~H. Golub.
\newblock Numerical methods for computing angles between linear subspaces.
\newblock \emph{Mathematics of Computation}, 27\penalty0 (123):\penalty0
  579--594, 1973.

\bibitem[Bollerslev et~al.(2009)Bollerslev, Tauchen, and
  Zhou]{bollerslev2009expected}
Tim Bollerslev, George Tauchen, and Hao Zhou.
\newblock Expected stock returns and variance risk premia.
\newblock \emph{Review of Financial Studies}, 22\penalty0 (11):\penalty0
  4463--4492, 2009.

\bibitem[Bollerslev et~al.(2020)Bollerslev, Li, Patton, and
  Quaedvlieg]{bollerslev2020semicov}
Tim Bollerslev, Jia Li, Andrew~J. Patton, and Rogier Quaedvlieg.
\newblock Realized semicovariances.
\newblock \emph{Econometrica}, 88\penalty0 (4):\penalty0 1515--1551, 2020.

\bibitem[Bonanno et~al.(2001)Bonanno, Lillo, and Mantegna]{bonanno2001hf}
Giovanni Bonanno, Fabrizio Lillo, and Rosario~N. Mantegna.
\newblock High-frequency cross-correlation in a set of stocks.
\newblock \emph{Quantitative Finance}, 1\penalty0 (1):\penalty0 96--104, 2001.

\bibitem[Britten-Jones and Neuberger(2000)]{brittenjones2000}
Mark Britten-Jones and Anthony Neuberger.
\newblock Option prices, implied price processes, and stochastic volatility.
\newblock \emph{Journal of Finance}, 55\penalty0 (2):\penalty0 839--866, 2000.

\bibitem[Buraschi et~al.(2014)Buraschi, Kosowski, and Trojani]{buraschi2014}
Andrea Buraschi, Robert Kosowski, and Fabio Trojani.
\newblock When there is no place to hide: Correlation risk and the
  cross-section of hedge fund returns.
\newblock \emph{Review of Financial Studies}, 27\penalty0 (2):\penalty0
  581--616, 2014.

\bibitem[Buss et~al.(2019)Buss, Sch{\"o}nleber, and Vilkov]{buss2019expected}
Adrian Buss, Lorenzo Sch{\"o}nleber, and Grigory Vilkov.
\newblock Expected correlation and future market returns.
\newblock Technical report, CEPR Discussion Paper No.\ 12760, 2019.
\newblock Working paper; SSRN 3114063.

\bibitem[Carr and Wu(2009)]{carr2009variance}
Peter Carr and Liuren Wu.
\newblock Variance risk premiums.
\newblock \emph{Review of Financial Studies}, 22\penalty0 (3):\penalty0
  1311--1341, 2009.

\bibitem[Carvalho and Ara{\'u}jo(2024)]{carvalho2024etf}
Lucas Carvalho and Tanya Ara{\'u}jo.
\newblock The dynamics of exchange traded funds: A geometrical and topological
  approach.
\newblock \emph{Applied Network Science}, 9:\penalty0 73, 2024.
\newblock \doi{10.1007/s41109-024-00674-8}.

\bibitem[{Cboe}(2019)]{cboe2019vix}
{Cboe}.
\newblock {Cboe} volatility index ({VIX}) white paper.
\newblock Technical report, Chicago Board Options Exchange, 2019.

\bibitem[Collin-Dufresne and Goldstein(2002)]{collindufresne2002span}
Pierre Collin-Dufresne and Robert~S. Goldstein.
\newblock Do bonds span the fixed income markets? theory and evidence for
  unspanned stochastic volatility.
\newblock \emph{Journal of Finance}, 57\penalty0 (4):\penalty0 1685--1730,
  2002.

\bibitem[Corsi(2009)]{corsi2009har}
Fulvio Corsi.
\newblock A simple approximate long-memory model of realized volatility.
\newblock \emph{Journal of Financial Econometrics}, 7\penalty0 (2):\penalty0
  174--196, 2009.

\bibitem[Davis and Kahan(1970)]{davis1970rotation}
Chandler Davis and W.~M. Kahan.
\newblock The rotation of eigenvectors by a perturbation. {III}.
\newblock \emph{SIAM Journal on Numerical Analysis}, 7\penalty0 (1):\penalty0
  1--46, 1970.

\bibitem[Demeterfi et~al.(1999)Demeterfi, Derman, Kamal, and
  Zou]{demeterfi1999}
Kresimir Demeterfi, Emanuel Derman, Michael Kamal, and Joseph Zou.
\newblock More than you ever wanted to know about volatility swaps.
\newblock Technical report, Goldman Sachs Quantitative Strategies Research
  Notes, 1999.

\bibitem[Dew-Becker and Giglio(2016)]{dewbecker2016frequency}
Ian Dew-Becker and Stefano Giglio.
\newblock Asset pricing in the frequency domain: Theory and empirics.
\newblock \emph{Review of Financial Studies}, 29\penalty0 (8):\penalty0
  2029--2068, 2016.

\bibitem[Driessen et~al.(2009)Driessen, Maenhout, and
  Vilkov]{driessen2009price}
Joost Driessen, Pascal~J. Maenhout, and Grigory Vilkov.
\newblock The price of correlation risk: Evidence from equity options.
\newblock \emph{Journal of Finance}, 64\penalty0 (3):\penalty0 1377--1406,
  2009.

\bibitem[Eleut{\'e}rio et~al.(2014)Eleut{\'e}rio, Ara{\'u}jo, and
  Vilela~Mendes]{eleuterio2014portfolios}
S.~Eleut{\'e}rio, Tanya Ara{\'u}jo, and R.~Vilela~Mendes.
\newblock Portfolios and the market geometry.
\newblock \emph{Physica A: Statistical Mechanics and its Applications},
  410:\penalty0 226--235, 2014.
\newblock \doi{10.1016/j.physa.2014.05.016}.

\bibitem[Epps(1979)]{epps1979}
Thomas~W. Epps.
\newblock Comovements in stock prices in the very short run.
\newblock \emph{Journal of the American Statistical Association}, 74\penalty0
  (366):\penalty0 291--298, 1979.

\bibitem[George et~al.(2023)George, Kachhara, and Ambika]{george2023}
Sandip~V. George, Sneha Kachhara, and G.~Ambika.
\newblock Early warning signals for critical transitions in complex systems.
\newblock \emph{Physica Scripta}, 98\penalty0 (7):\penalty0 072002, 2023.
\newblock \doi{10.1088/1402-4896/acde20}.

\bibitem[Halperin(2026{\natexlab{a}})]{halperin2026a}
Igor Halperin.
\newblock Observable matrix dynamics of stocks.
\newblock \emph{arXiv preprint arXiv:2607.19005 [q-fin.ST]},
  2026{\natexlab{a}}.
\newblock Code: \url{https://github.com/ighalp/omd_finance}.

\bibitem[Halperin(2026{\natexlab{b}})]{halperin2026b}
Igor Halperin.
\newblock Are three matrices all you need to beat the market? observable matrix
  dynamics for portfolio optimization.
\newblock \emph{arXiv preprint arXiv:2607.27461 [q-fin.PM]},
  2026{\natexlab{b}}.

\bibitem[Hansen and Hodrick(1980)]{hansen1980forward}
Lars~Peter Hansen and Robert~J. Hodrick.
\newblock Forward exchange rates as optimal predictors of future spot rates: An
  econometric analysis.
\newblock \emph{Journal of Political Economy}, 88\penalty0 (5):\penalty0
  829--853, 1980.

\bibitem[Kritzman et~al.(2011)Kritzman, Li, Page, and Rigobon]{kritzman2011pca}
Mark Kritzman, Yuanzhen Li, Sebastien Page, and Roberto Rigobon.
\newblock Principal components as a measure of systemic risk.
\newblock \emph{Journal of Portfolio Management}, 37\penalty0 (4):\penalty0
  112--126, 2011.

\bibitem[Laloux et~al.(1999)Laloux, Cizeau, Bouchaud, and
  Potters]{laloux1999noise}
Laurent Laloux, Pierre Cizeau, Jean-Philippe Bouchaud, and Marc Potters.
\newblock Noise dressing of financial correlation matrices.
\newblock \emph{Physical Review Letters}, 83\penalty0 (7):\penalty0 1467--1470,
  1999.

\bibitem[Mar{\v{c}}enko and Pastur(1967)]{marchenko1967}
Vladimir~A. Mar{\v{c}}enko and Leonid~A. Pastur.
\newblock Distribution of eigenvalues for some sets of random matrices.
\newblock \emph{Mathematics of the USSR-Sbornik}, 1\penalty0 (4):\penalty0
  457--483, 1967.

\bibitem[M{\"u}ller et~al.(1997)M{\"u}ller, Dacorogna, Dav{\'e}, Olsen, Pictet,
  and von Weizs{\"a}cker]{muller1997volatilities}
Ulrich~A. M{\"u}ller, Michel~M. Dacorogna, Rakhal~D. Dav{\'e}, Richard~B.
  Olsen, Olivier~V. Pictet, and Jakob~E. von Weizs{\"a}cker.
\newblock Volatilities of different time resolutions --- analyzing the dynamics
  of market components.
\newblock \emph{Journal of Empirical Finance}, 4\penalty0 (2--3):\penalty0
  213--239, 1997.

\bibitem[Newey and West(1987)]{newey1987}
Whitney~K. Newey and Kenneth~D. West.
\newblock A simple, positive semi-definite, heteroskedasticity and
  autocorrelation consistent covariance matrix.
\newblock \emph{Econometrica}, 55\penalty0 (3):\penalty0 703--708, 1987.

\bibitem[Onnela et~al.(2003)Onnela, Chakraborti, Kaski, Kert{\'e}sz, and
  Kanto]{onnela2003dynamics}
J.-P. Onnela, A.~Chakraborti, K.~Kaski, J.~Kert{\'e}sz, and A.~Kanto.
\newblock Dynamics of market correlations: Taxonomy and portfolio analysis.
\newblock \emph{Physical Review E}, 68:\penalty0 056110, 2003.

\bibitem[Plerou et~al.(2002)Plerou, Gopikrishnan, Rosenow, Amaral, Guhr, and
  Stanley]{plerou2002rmt}
Vasiliki Plerou, Parameswaran Gopikrishnan, Bernd Rosenow, Lu{\'i}s A.~Nunes
  Amaral, Thomas Guhr, and H.~Eugene Stanley.
\newblock Random matrix approach to cross correlations in financial data.
\newblock \emph{Physical Review E}, 65:\penalty0 066126, 2002.

\bibitem[Politis and Romano(1994)]{politis1994stationary}
Dimitris~N. Politis and Joseph~P. Romano.
\newblock The stationary bootstrap.
\newblock \emph{Journal of the American Statistical Association}, 89\penalty0
  (428):\penalty0 1303--1313, 1994.

\bibitem[Quax et~al.(2013)Quax, Kandhai, and Sloot]{quax2013}
Rick Quax, Drona Kandhai, and Peter M.~A. Sloot.
\newblock Information dissipation as an early-warning signal for the {Lehman
  Brothers} collapse in financial time series.
\newblock \emph{Scientific Reports}, 3:\penalty0 1898, 2013.
\newblock \doi{10.1038/srep01898}.

\bibitem[Stambaugh(1999)]{stambaugh1999}
Robert~F. Stambaugh.
\newblock Predictive regressions.
\newblock \emph{Journal of Financial Economics}, 54\penalty0 (3):\penalty0
  375--421, 1999.

\bibitem[Valkanov(2003)]{valkanov2003}
Rossen Valkanov.
\newblock Long-horizon regressions: Theoretical results and applications.
\newblock \emph{Journal of Financial Economics}, 68\penalty0 (2):\penalty0
  201--232, 2003.

\bibitem[Vilela~Mendes et~al.(2002)Vilela~Mendes, Lima, and
  Ara{\'u}jo]{vilelamendes2002process}
R.~Vilela~Mendes, R.~Lima, and Tanya Ara{\'u}jo.
\newblock A process-reconstruction analysis of market fluctuations.
\newblock \emph{International Journal of Theoretical and Applied Finance},
  5\penalty0 (8):\penalty0 797--821, 2002.

\bibitem[Vilela~Mendes et~al.(2003)Vilela~Mendes, Ara{\'u}jo, and
  Lou{\c{c}}{\~a}]{vilelamendes2003reconstructing}
R.~Vilela~Mendes, Tanya Ara{\'u}jo, and Francisco Lou{\c{c}}{\~a}.
\newblock Reconstructing an economic space from a market metric.
\newblock \emph{Physica A}, 323:\penalty0 635--650, 2003.

\bibitem[Whaley(1993)]{whaley1993}
Robert~E. Whaley.
\newblock Derivatives on market volatility: Hedging tools long overdue.
\newblock \emph{Journal of Derivatives}, 1\penalty0 (1):\penalty0 71--84, 1993.

\bibitem[Yu et~al.(2015)Yu, Wang, and Samworth]{yu2015davis}
Yi~Yu, Tengyao Wang, and Richard~J. Samworth.
\newblock A useful variant of the {Davis--Kahan} theorem for statisticians.
\newblock \emph{Biometrika}, 102\penalty0 (2):\penalty0 315--323, 2015.

\bibitem[Zlotnikov et~al.(2024)Zlotnikov, Liu, Halperin, He, and
  Huang]{zlotnikov2024}
Vadim Zlotnikov, Jiahao Liu, Igor Halperin, Fanyu He, and Lisa~L. Huang.
\newblock Model-free market risk hedging using crowding networks.
\newblock \emph{Journal of Portfolio Management}, 50\penalty0 (9):\penalty0
  132--141, 2024.

\end{thebibliography}
